\documentclass[pra,onecolumn,floatfix,superscriptaddress,longbibliography,notitlepage, nofootinbib]{revtex4-1}
\pdfoutput=1

\usepackage[utf8]{inputenc}
\usepackage{braket}
\usepackage{amsmath}
\DeclareMathOperator{\Tr}{Tr}
\usepackage{dcolumn}   
\usepackage{bm}        
\usepackage{amssymb}
\usepackage{mathtools}
\usepackage{tikz}
\usepackage{qcircuit}
\usepackage{appendix}
\usepackage{hyperref}
\usepackage{subcaption}
\usepackage{amsmath,amsfonts,amssymb}
\usepackage{mathtools}
\usepackage{thmtools,thm-restate}
\usepackage{algorithm}
\usepackage{mathrsfs}
\usepackage{pgfplots}
\usetikzlibrary{3d}
\usepackage{xy}

\usepackage{algpseudocode}

\newtheorem{definition}{Definition}
\newtheorem{theorem}{Theorem}
\newtheorem{lemma}{Lemma}

\newtheorem{method}{Algorithm}

\newcommand{\revise}[1]{\textcolor{black}{#1}}
\newcommand{\xbf}{\textbf{x}}

\newcommand{\Ibb}{\mathbb{I}}

\newcommand{\Rbb}{\mathbb{R}}

\newcommand{\Fsr}{\mathscr{F}}
\newcommand{\Jcal}{\mathcal{J}}

\usetikzlibrary{arrows.meta}

\def\BibTeX{{\rm B\kern-.05em{\sc i\kern-.025em b}\kern-.08em
    T\kern-.1667em\lower.7ex\hbox{E}\kern-.125emX}}

\begin{document}
\title{Quantum Speedup in Dissecting Roots and Solving Nonlinear Algebraic Equations }
\author{Nhat A. Nghiem}
\affiliation{Department of Physics and Astronomy, State University of New York at Stony Brook, Stony Brook, NY 11794-3800, USA}
\affiliation{C. N. Yang Institute for Theoretical Physics, State University of New York at Stony Brook, Stony Brook, NY 11794-3840, USA}

\begin{abstract}
   It is shown that quantum computer can detect the existence of root of a function almost exponentially more efficient than the classical counterpart. It is also shown that a quantum computer can produce quantum state corresponding to the solution of nonlinear algebraic equations quadratically faster than the best known classical approach. Various applications and implications are discussed, including a quantum algorithm for solving dense linear systems with quadratic speedup, determining equilibrium states, simulating the dynamics of nonlinear coupled oscillators, estimating Lyapunov exponent, an improved quantum partial differential equation solver, and a quantum-enhanced collision detector for robotic motion planning. This provides further evidence of quantum advantage without requiring coherent quantum access to classical data, delivering meaningful real-world applications.
\end{abstract}
\maketitle

\section{Introduction}
Quantum mechanics has revolutionized our understanding of the physical world, revealing that microscopic properties can be fundamentally distinct from macroscopic ones. This distinction has inspired a new computational paradigm in which quantum properties are harnessed for information processing. Beyond its impact on physics, quantum mechanics holds the potential to revolutionize technology, a field that has traditionally relied on digital computation. The concept of quantum computing was first proposed in early works \cite{manin1980computable, benioff1980computer}. Subsequent breakthroughs \cite{shor1999polynomial, grover1996fast, deutsch1985quantum, deutsch1992rapid} demonstrated that by leveraging intrinsic quantum effects such as entanglement and superposition, quantum computers can factorize integers, search unstructured databases, and analyze functions implemented by black-box models faster than any known classical methods. These pioneering works laid the foundation for quantum computing and have had a lasting impact on its development, inspiring numerous subsequent advancements \cite{grover1996fast, shor1999polynomial, feynman2018simulating, lloyd1996universal, berry2007efficient, berry2012black, berry2014high, berry2015hamiltonian, low2017optimal, low2019hamiltonian, childs2022quantum, o2016scalable, cerezo2021variational, babbush2023exponential, babbush2018low, mitarai2023perturbation, robert2021resource, kitaev1995quantum, aharonov2006polynomial, childs2010relationship, childs2021high, childs2017quantum, childs2021quantum, lloyd2013quantum, lloyd2014quantum, lloyd2016quantum, lloyd2020quantum, hauke2020perspectives, liu2021efficient, liu2018quantum, liu2024towards, durr1996quantum, deutsch1992rapid, bauer2020quantum, brassard1997quantum, brassard2002quantum, jordan2012quantum, leyton2008quantum, nachman2021quantum, jordan2005fast, garnerone2012adiabatic, gilyen2022quantum, miessen2023quantum, haah2021quantum, arrazola2019quantum, kerenidis2019quantum, hallgren2007polynomial}.

The central objective of the field is to demonstrate quantum advantage—a scenario where quantum computers outperform classical ones in solving computational problems. Despite many successes, most claimed quantum speed-ups remain unjustified, as they rely on an oracle or black-box model that assumes quantum computers can access classical data in a coherent quantum manner. This assumption presents a significant challenge to the practical applicability of quantum algorithms. Recent studies \cite{tang2018quantum,tang2019quantum,tang2021quantum} have even shown that if classical computers are provided with an analogous technique, known as $l_2$-sampling, they can perform tasks such as supervised learning and principal component analysis with at most a polynomial slowdown compared to quantum methods. Consequently, many previously claimed exponential quantum speed-ups have been refuted. More critical issues surrounding quantum algorithms are discussed in \cite{aaronson2015read}, raising the fundamental question of whether quantum advantage can be achieved in a meaningful and practical way.

Nonetheless, there exist a few rigorous demonstrations of quantum advantage. Bravyi et al.\cite{bravyi2018quantum} showed that quantum computers can solve certain linear algebraic problems involving binary quadratic forms with constant depth, and this advantage persists even in the presence of noise \cite{bravyi2020quantum}. Maslov et al.\cite{maslov2021quantum} demonstrated that in a space-restricted model, quantum computers outperform classical ones. Liu et al.~\cite{liu2021rigorous} identified a supervised learning problem where a quantum computer can achieve efficient performance, whereas a classical computer performs no better than random guessing. While these results are theoretically significant, they remain largely impractical. Most recently, \cite{nghiem2024simple, nghiem2025quantum} demonstrated that quantum computers do not necessarily require coherent quantum access to perform gradient descent. This insight has led to a variety of new quantum algorithms, including quantum linear system solvers, least-square data fitting algorithms, support vector machines, supervised learning models, quantum neural network training methods, and quantum algorithms for determining ground and excited state energies. These developments suggest new prospects for quantum computational advantage and its applications.

In this work, we ride on the success established in \cite{nghiem2024simple} and \cite{nghiem2025quantum} by investigating analytical function properties in an oracle-free regime. Specifically, we focus on two fundamental problems: the existence of roots and the determination of roots for polynomials. For the first problem, we demonstrate that given knowledge of a function (e.g., polynomial coefficients), a quantum computer can determine whether a root—where the function value is zero—exists within a given domain. Our method is based on the insight that the existence of a root implies a change in the function's sign. By examining sign changes at multiple points within the domain, we can efficiently detect the presence of a root. The ability of quantum computers to process multiple points simultaneously with logarithmic resources enables us to determine root existence in logarithmic time. Furthermore, by establishing a lower bound for classical algorithms, which requires at least linear time in the number of points, we achieve an almost exponential quantum speed-up. For the second problem, we develop a quantum Newton’s method, which is a direct quantum counterpart of the classical Newton’s method for solving systems of nonlinear algebraic equations. We prove that the quantum state corresponding to the solution of a given nonlinear system can be obtained in quadratic time relative to the system's dimension, yielding a quadratic speed-up over classical approaches. To highlight the potential applications of our results, we extend and apply our methods to various contexts. A direct application of quantum Newton’s method results in a dense linear system solver with quadratic speed-up. Additionally, we analyze the dynamics of nonlinear coupled oscillators, leveraging our quantum Newton’s method to find equilibrium states and simulate their behavior over time. An unexpected yet significant corollary of our techniques is the development of an efficient circulant matrix construction method, which improves upon the quantum algorithm for solving partial differential equations proposed in \cite{childs2021high}. Finally, we explore a potential application in robotic motion planning, where our quantum root-dissection algorithm can be used for collision detection.

Before delving into the technical details, we outline the structure of our paper. We first provide in Sec.~\ref{sec: overview} a formal overview, related assumptions, technical summaries and statement of main results, including Theorem \ref{thm: univariateroot}, Theorem \ref{thm: univariateroot} and Theorem \ref{thm: quantumnewton}. A more detailed version of quantum root dissecting algorithm is outlined in Sec.~\ref{sec: quantumdissecting}, followed by a detailed analysis and discussion is included. Section \ref{sec: quantumnewtonmethod} is devoted for the details of quantum algorithm for solving nonlinear algebraic equations, followed by a thorough discussion on complexity and possible extension in Section \ref{sec: disscusionandgeneralizationnewtonmethod}. Applications and implications of our two main quantum algorithms are provided in Section \ref{sec: application}. 

\section{Overview and Summary of Results}
\label{sec: overview}
As mentioned, this section aims at providing a formal description of our main objectives, including dissecting the existence of a root within a known domain, and solving nonlinear algebraic equations. In addition, we also provide key information and related assumptions for our subsequent algorithms. To better convey the main spirit of this work, we also provide a technical summary of our two main quantum algorithms, and statement of main results in Theorem \ref{thm: univariateroot}, \ref{thm: multivariateroot}, \ref{thm: quantumnewton}. 

\subsection{The first problem: dissecting roots within a domain}
A root of a univariate function $f(x)$ is a point where $f(x)=0$. In principle, if we know that there exists a root within the given domain $\mathscr{D}$, then numerical methods such as Newton-Graphson or Levenberg-Marquardt algorithm can be applied to find the root. 
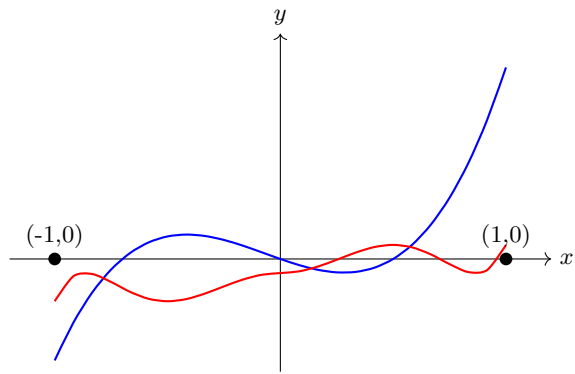
\begin{figure}[h!]
    \centering
    \begin{tikzpicture}[scale = 3.0]
        \draw[->] (-1.2, 0) -- (1.2, 0) node[right] {$x$};
    \draw[->] (0, -0.5) -- (0, 1) node[above] {$y$};
    \draw[domain = -1:1, smooth, variable = \x, blue, thick] plot( {\x}, {(\x*(\x-0.5)*(\x+0.7)} );
    \draw[domain = -1:1, smooth, variable = \x, red, thick] plot( {\x}, { (1/16)*( 32*\x^6 - 48*\x^4 +18*\x^2 -1)  }  );
    \fill[black] (-1,0) circle (0.8pt);
    \fill[black] (1,0) circle (0.8 pt);
    \node at (1, 0.1) [black] {(1,0)} ;
    \node at (-1, 0.1) [black] {(-1,0)}; 
    \end{tikzpicture}
    \caption{\textit{The plot of polynomial $f(x)= x(x-0.5)(x+0.7)$ and $g(x) = 2x^6 - 3x^4 +  \frac{9}{8}x^2 - \frac{1}{16} $ in the domain $[-1,1]$. The root of $f(x), g(x)$ is the intersection of the graph of $f(x), g(x)$ with the $x$-axis. }  }
    \label{fig: root}
\end{figure} 
Our objective is to dissect whether or not a given function has at least one root within some domain. Without loss of generality, let the domain $\mathscr{D} \subset [-\frac{1}{2},\frac{1}{2}]$. Furthermore, we assume that the function of consideration $f(x)$ is a continuous polynomial or admits a polynomial approximation within $\mathscr{D}$. In principle, any analytical function can be (locally) approximated with polynomials by virtue of Taylor's expansion. In addition, we assume that  $f(x)$ is bounded as $|f(x)| \leq \frac{1}{2}$ in the domain $\mathscr{D}$. We point out the following fundamental property of a function, which is called \textit{intermediate value theorem}:

\noindent
\textbf{Criterion: } \textit{If $f(x)$ is continuous on $\mathscr{D}$, and $\rm sign \ \big(f(x_1)f(x_2)\big) < 0$ for $x_1,x_2 \in \mathscr{D}$, then there is at least one root in $[x_1,x_2]$. }

Our (soon to be outlined) quantum algorithm for dissecting root of $f(x)$ is built upon an extension of the above criterion. The intuition is that, the existence of root indicates that the function intersects with $x$-axis, implying that the sign of function has changed. If we look at, say, $n$ points $x_1,x_2,...x,_n$, then if the value $f(x_i)$ for some $1 \leq i \leq n$ is less than zero and for another $j$, $f(x_j)$ is greater than zero, it means that there is at least one point of intersection, thus implying the root. Given $n$ points, $x_1,x_2,...x,_n$, there are $n$ values $f(x_1),f(x_2),...,f(x_n)$, if we evaluate all the values one by one, then it will take $n$ time. Certainly this strategy will not provide any quantum advantage. Instead, we look at $\min \{ f(x_1),f(x_2),...,f(x_n) \}$ and $\max \{ f(x_1),f(x_2),...,f(x_n)\}$, if there is any change of sign, it indicates that there is at least one root.

A key technical aspect of our subsequent algorithms is based on the following. Instead of working with vector/state convention, we embed it, say a vector $\xbf =\sum_{j=1}^n x_j\ket{j-1} $, to a diagonal operator $\rm diag (\xbf)$ that contains $\xbf$ on the diagonal, i.e.,
\begin{align}
   \sum_{j=1}^n x_j\ket{j-1} = \begin{pmatrix}
        x_1\\
        x_2 \\
        \vdots \\
        x_n
    \end{pmatrix} \longrightarrow  \bigoplus_{j=1}^n x_j=  \begin{pmatrix}
        x_1 & 0 & 0 & 0 \\
        0 & x_2 & 0 & 0 \\
        0 & 0 & \ddots & 0 \\
        0 & 0 & 0 & x_n
    \end{pmatrix} 
\end{align}
We mention a few simple algebraic properties of diagonal operator. For $\rm diag(\xbf) = \bigoplus_j x_j $ and $\rm diag (\textbf{y}) = \bigoplus_j y_j$, we have $\rm diag(\xbf) + \rm diag( \textbf{y}) = \bigoplus_j (x_j + y_j) $ and $ \rm diag(\xbf)  \cdot \rm diag( \textbf{y}) = \bigoplus_j x_jy_j$. As we shall see, by changing the working regime from a conventional vector-base to diagonal operator representation, we can utilize many arithmetic recipes from the versatile block-encoding, or quantum singular value transformation framework \cite{gilyen2019quantum}. Generally speaking, a unitary $U$ is said to be a block encoding of $A$ (with operator norm $|A| \leq 1$) if $U$ contains $A$ in the top left corner, that is, $ U = \begin{pmatrix}
        A & \cdot \\
        \cdot & \cdot 
    \end{pmatrix}$, where $(\cdot)$ refers to possibly non-zero entries. Equivalently, $A =( \bra{\bf 0}\otimes \Ibb )  \ U  \ (\ket{\bf 0}\otimes \Ibb)$ where $\ket{\bf 0}$ is the ancillary qubits required for the block encoding purpose. Such a block encoding allows us to arithmetically manipulate the block-encoded operator easily.  Suppose that $U_1$ is a block encoding of $A_1$, $U_2$ is a block encoding of $A_2$, then for some known $\alpha, \beta \leq 1$, we can construct another unitary a unitary block encoding of $\alpha_1 A_1, \alpha_2 A_2$ (Lemma \ref{lemma: scale}, trivial scaling) $ \alpha A_1 + \beta A_2$ (Lemma \ref{lemma: sumencoding}, Linear combination), of $ A_1 A_2$ (Lemma \ref{lemma: product}, Multiplication), and also of $A_1 \otimes A_2$ (Ref. \cite{camps2020approximate}, Tensor product). Additionally, for a factor $\gamma > 1$ and with a guarantee $\gamma A \leq \frac{1}{2}$, it is possible to construct the block encoding of $\gamma A$ (Lemma \ref{lemma: amp_amp}, Amplification). We remark that more details are summarized in the Appendix \ref{sec: prelim}, and we highly encourage the readers to take a glimpse over Appendix \ref{sec: prelim}. 
With the above notations, the following diagram summarizes the flow and key steps of our quantum algorithm for dissecting existence of roots: 
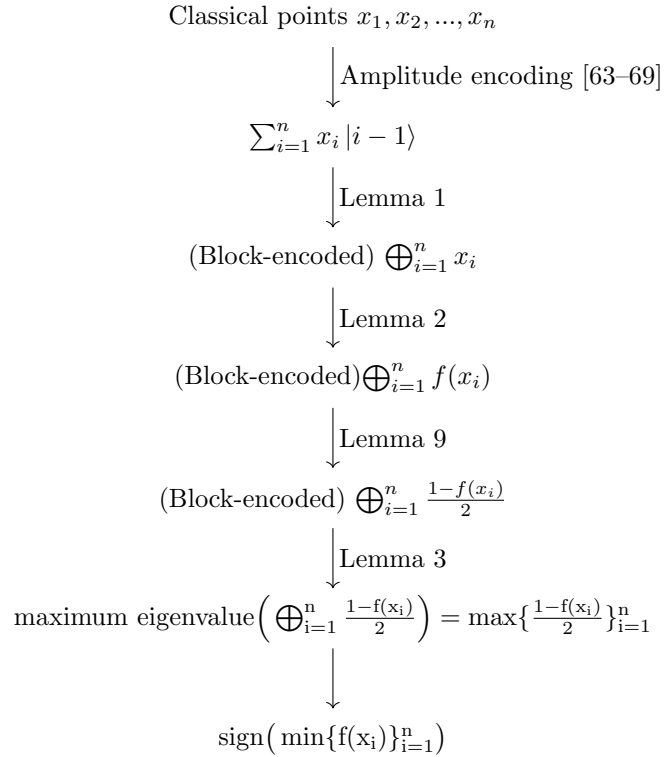
\begin{center}
    \begin{tikzpicture}[scale = 0.8]
        \node at (0,5) {Classical points $x_1,x_2,...,x_n$}; 
        \draw[->] (0,4.5) -- (0,3.5); 
        \node at (2.8, 4) {Amplitude encoding \cite{grover2000synthesis,grover2002creating,plesch2011quantum, schuld2018supervised, nakaji2022approximate,marin2023quantum,zoufal2019quantum} };
        \node at (0,3) { $\sum_{i=1}^n x_i \ket{i-1}$ };
        \draw[->] (0,2.5) -- (0,1.5); 
        \node at (1, 2) {Lemma \ref{lemma: diagonal}  }; 
        \node at (0,1) {(Block-encoded) $\bigoplus_{i=1}^n x_i$ } ;
        \draw[->] (0,0.5) -- (0,-0.5); 
        \node at (1,0) {Lemma \ref{lemma: theorem56}};
        \node at (0,-1) { (Block-encoded)$\bigoplus_{i=1}^n f(x_i)$ }; 
        \draw[->] (0,-1.5) -- (0,-2.5); 
        \node at (1,-2) {Lemma \ref{lemma: sumencoding}};
        \node at (0,-3) { (Block-encoded) $\bigoplus_{i=1}^n \frac{1-f(x_i)}{2}$  }; 
        \draw[->] (0,-3.5) -- (0,-4.5); 
        \node at (1,-4) {Lemma \ref{lemma: largestsmallest}};
        \node at (0,-5) {$ \rm maximum \ eigenvalue\Big(\bigoplus_{i=1}^n \frac{1-f(x_i)}{2} \Big)  =  \max \{ \frac{1-f(x_i)}{2} \}_{i=1}^n  $}; 
        \node at (0,-7) { $ \rm sign \big( \min \{ f(x_i)  \}_{i=1}^n  \big)$};
        \draw[->] (0,-5.5) -- (0,-6.5) ;
    \end{tikzpicture}
\end{center}
To this end, we recapitulate the result of the algorithm above in the following:
\begin{theorem}
\label{thm: univariateroot}
    Given a univariate function $f(x)$ defined as above. Let the domain of working $\mathscr{D} \subset [-\frac{1}{2}, \frac{1}{2}]$ and let $x_1,x_2,...x,_n \in \mathscr{D}$ be $n$ points of consideration.  Quantum algorithm can reveals if there is a change of sign among $f(x_1),f(x_2),...,f(x_n)$ with complexity $\mathcal{O}\big( \log^2 n \big)$, meanwhile classical algorithm requires $\Omega(n)$.  
\end{theorem}
In particular, the above strategy and result can be extended to higher dimension, yielding the following:
\begin{theorem}
\label{thm: multivariateroot}
    Given a $M$-variable function $f(\xbf)$. Let the domain of working $\mathscr{D} \subset [-1,1]^M$ and let $\xbf_1,\xbf_2,...,\xbf_n \in \mathscr{D}$ be $n$ points of consideration.  Quantum algorithm can reveals if there is a change of sign among $f(\xbf_1),f(\xbf_2),...,f(\xbf_n)$ with complexity $\mathcal{O}\big( \log^2 n \big)$, meanwhile classical algorithm requires $\Omega(n)$.  
\end{theorem}
The above theorem indicates that quantum computer can dissect the sign of function $f(x)$ almost exponentially faster than classical one. Thus, it clearly implies that there is quantum advantage in the task of determining if the given function $f(x)$ has at least one root. As fueled by the above success, we consider another related problem, that is, finding the root, given that we already determine its existence.

\subsection{The second problem: nonlinear algebraic equations}
\label{sec: overviewnonlinear}
A system of equations with $n$ variables is:
\begin{align}
   \Fsr(\xbf) = \begin{cases}
        f_1(\xbf)   \\
        f_2(\xbf)   \\
        \vdots \\
        f_n(\xbf) 
    \end{cases}
    \label{nonlinearsystem}
\end{align}
where $\xbf = (x_1,x_2,...,x_n)$ is a $n$-dimensional vector and each $f_j(\xbf):\Rbb^n \longrightarrow \Rbb$ is a $n$-variable function. The goal is to find $\xbf$ such that $\Fsr(\xbf) = \Vec{0}$, or equivalently, for all $j=1,2,...,n$, $f_j(\xbf) = 0$. If all $f_j(\xbf)$ are linear functions of $\xbf$, then the above system is called linear algebraic equations. Otherwise, for $f_j(\xbf)$ to be non-linear, the system is non-linear algebraic equations. We focus on the latter case.
\begin{figure}[H]
    \centering
    \begin{subfigure}[b]{0.48\textwidth}
    \centering
        \begin{tikzpicture}
        \begin{axis}[
            enlargelimits,
            xlabel={$x$},
            ylabel={$y$},
            grid=major,
            legend pos=north east,
            samples=100
        ]
        \addplot[
            domain=-2:2,
            thick,
            red
        ]
        {1 - x^2}; 
        \addlegendentry{$f_1(x,y) = 0$}
        
        \addplot[
            domain=-2:2,
            thick,
            blue
        ]
        {exp(x) - 1}; 
        \addlegendentry{$f_2(x,y) = 0$}
        
        \addplot[only marks, mark=*, mark size=2pt, black] coordinates {
            (0,0) 
        };
        \node[above right] at (axis cs:0,0) {\small $(x^*, y^*)$};
        
        \end{axis}
    \end{tikzpicture}
    \caption{The plot of two function $f_1(x,y) = y+x^2-1$ and $f_2(x,y) = y-e^x +1$ }
    \label{fig: fig1}
    \end{subfigure}
    \hfill
    \begin{subfigure}[b]{0.5\textwidth}
        \centering
    \begin{tikzpicture}
        \begin{axis}[
            xlabel={$x$},
            ylabel={$y$},
            zlabel={$z$},
            view={135}{30}, 
            grid=major,
            legend pos=north east
        ]
        
        \addplot3[
            surf,
            opacity=0.5,
            domain=-2:2,
            y domain=-2:2,
            colormap/redyellow
        ]
        {x^2 + y^2 - 1}; 
        
        \addlegendentry{$f_1(x,y,z) = 0$}

        \addplot3[
            surf,
            opacity=0.5,
            domain=-2:2,
            y domain=-2:2,
            colormap/bluered
        ]
        {sin(deg(x)) + cos(deg(y))}; 
        
        \addlegendentry{$f_2(x,y,z) = 0$}

        \addplot3[
            surf,
            opacity=0.5,
            domain=-2:2,
            y domain=-2:2,
            colormap/viridis
        ]
        {0.5*x + 0.5*y}; 
        
        \addlegendentry{$f_3(x,y,z) = 0$}

        \addplot3[only marks, mark=*, mark size=2pt, black] coordinates {
            (0,0,0)
        };
        \node[above] at (axis cs:0,0,0) {\small $(x^*, y^*, z^*)$};

        \end{axis}
    \end{tikzpicture}

    \caption{Plot of three functions: $f_1(x,y,z) = z-x^2 -y^2 +1, f_2(x,y,z) = z- \sin(x) - \cos(y), f_3(x,y,z) = z- 0.5x-0.5y$}
    \label{fig: fig2}
    \end{subfigure}
    \caption{ \textit{Illustration of nonlinear system of equations and related problem of finding $\xbf$ such that it simultaneously satisfies all equations in 2-d and 3-d, respectively. In the language of algebraic geometry, each $f_j(\xbf) = 0 $ defines an algebraic variety. In 2-d, a variety defines a curve, meanwhile in 3-d, a variety defines a plane. Solving nonlinear algebraic equations is equivalent to finding the intersection of these varieties.}   }
\end{figure}
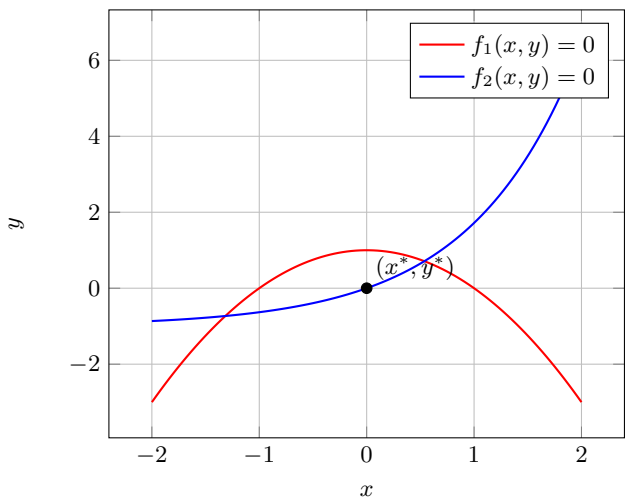
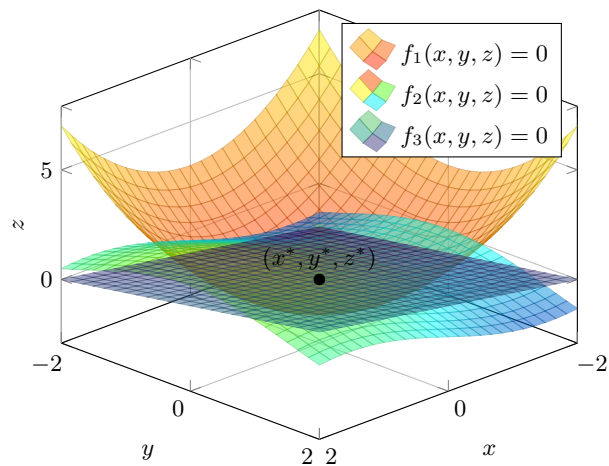
A popular method for solving nonlinear algebraic equations is Newton's method, which works as follows. We begin with some initial guess $\xbf_0$. Suppose without loss of generality that at $t$-th iteration step, the temporal solution is $\xbf_t$. Define the Jacobian at $t$-th iteration step as:
\begin{align}
    \Jcal(\xbf_t) = \begin{pmatrix}
        \cdots &  \bigtriangledown^T f_1(\xbf_t) &  \cdots \\
        \cdots &  \bigtriangledown^T f_2(\xbf_t) & \cdots \\
        \vdots  & \vdots  & \vdots  \\
        \cdots &  \bigtriangledown^T f_n(\xbf_t) & \cdots 
    \end{pmatrix}
\end{align}
That is, the $j$-th row of $J(\xbf_k)$ is the gradient of $f_j$ evaluated at $\xbf_t$. Define the following linear equation $\Jcal(\xbf_t) \Delta_t = \Fsr (\xbf_t)$. Once solving the equation, we update the temporal solution $\xbf_t$ as $\xbf_{t+1} = \xbf_t - \Delta_t$. Then the above procedure is iterated for a total of $T$ times (typically $T$ is user-dependent), or until convergence. Since the Newton's method begins with some random initialization $\xbf_0$, the outcome depends on initial condition as well. More specifically, the aforementioned iterative procedure gradually ``drives'' the initial guess $\xbf_0$ to the closest $\xbf$ that satisfies the nonlinear algebraic equations. Different initial beginnings would lead to different solutions, if there are more than one. 

Another approach for solving nonlinear algebraic equations is the Levenberg-Marquardt method, which is based on converting the original problem to an optimization problem, i.e., find $\xbf$ satisfying $\min_{\xbf} S(\xbf) = \frac{1}{2} \sum_{j=1}^n f_j^2(\xbf)$. The procedure for this approach is quite similar to Newton's method. We begin with an initial guess $\xbf_0$. At $t$-th step, let the Jacobian $\Jcal(\xbf_t)$ be as above. Define a linear system $\big( \Jcal(\xbf_t)^T \Jcal(\xbf_t) +\lambda \Ibb  \big) \Delta_t = -\Jcal^T(\xbf_t) \Fsr(\xbf_t)$ where $\lambda$ is called a damping factor.  Solving this system yields $\Delta_t$, which can be used to update $\xbf_t$ as $\xbf_{t+1} = \xbf_t + \Delta_t$. Thus, in principle, the technique that we shall outline below, which is for Newton's method, can be modified simply to execute Levenberg-Marquardt algorithm. 

In quantum setting, instead of obtaining $\xbf$, the target is obtaining a quantum state $\ket{\xbf}$ that corresponds to $\xbf$. Our quantum Newton method (soon to be outlined) is a direct translation of classical Newton's method above into a quantum setting. We note that there are a few alternative quantum algorithms for solving nonlinear algebraic equations,  \cite{qian2019quantum, xue2021quantum, xue2022quantum, nghiem2024quantumnonlinear}, however, they are either heuristic or require oracle access to classical information. We emphasize that we do not require such an access, and as we shall see, the speed-up is rigorously derived.  

Similar to the previous case, we embed a vector $\xbf$ into a diagonal operator $\rm diag(\xbf)$, which enables the application of block-encoding techniques. Second, let  $\mathscr{D} \subset [-\frac{1}{2},\frac{1}{2}]^n$ be the domain of working. Next, for all $j$, let $f_j(\xbf)$ be bounded by $|f_j(\xbf)| \leq \frac{1}{2} $, and let its gradient be also bounded by $| \bigtriangledown f_j(\xbf) |_{\infty} \leq M$ for some $M$, and for all $\xbf \in \mathscr{D}$. In addition, for all $j$, let $f_j(\xbf)$ possess the same structure as discussed in \cite{nghiem2024simple}, more concretely, $f_j(\xbf)$ could be any of the following types:
\begin{align}
\label{eqn: function}
    \sum_{i=1}^K a_{i,1} x_1^i + a_{i,2} x_2^i + \cdots + a_{i,n} x_n^i, \   \sum_{i=1}^K  \big( a_{i,1} x_1 + a_{i,2} x_2 + \cdots + a_{i,n} x_n \big)^i,\ \prod_{i=1}^K  \big( a_{i,1} x_1 + a_{i,2}x_2 + \cdots + a_{i,n}x_n  + b_i \big)^i 
\end{align}
where $a_{i,1},a_{i,2},...,a_{i,n}$ are known coefficients and $P$ is a polynomial. In addition, for all $f_j(\xbf)$, there is an efficient quantum circuit, e.g., amplitude encoding techniques \cite{grover2000synthesis,grover2002creating,plesch2011quantum, schuld2018supervised, nakaji2022approximate,marin2023quantum,zoufal2019quantum, zhang2022quantum}. Especially, the result in \cite{zhang2022quantum}, that generates the state $\sim (a_{i,1},a_{i,2},...,a_{i,n})^T$ or $\sim (\sqrt{a_{i,1}}, \sqrt{a_{i,2}},..., \sqrt{a_{i,n}}) ^T$ (could be up to a normalization). Last, we assume that the absolute value of the smallest singular value of Jacobian is lower bounded by a constant $\frac{1}{\Lambda}$ for all $\xbf \in \mathscr{D}$. 

To provide a technical summary of our quantum Newton's algorithm, we recall that Newton's method is iterative, so suppose that at $t$-th time step, we are provided with a block encoding of $\rm diag(\xbf_t)$ where $\xbf_t$ is the temporal solution at the indicated time step. The key steps and flow of our algorithm are summarized in the following diagram (note that all operators included in this diagram are block encoded by some unitary): 
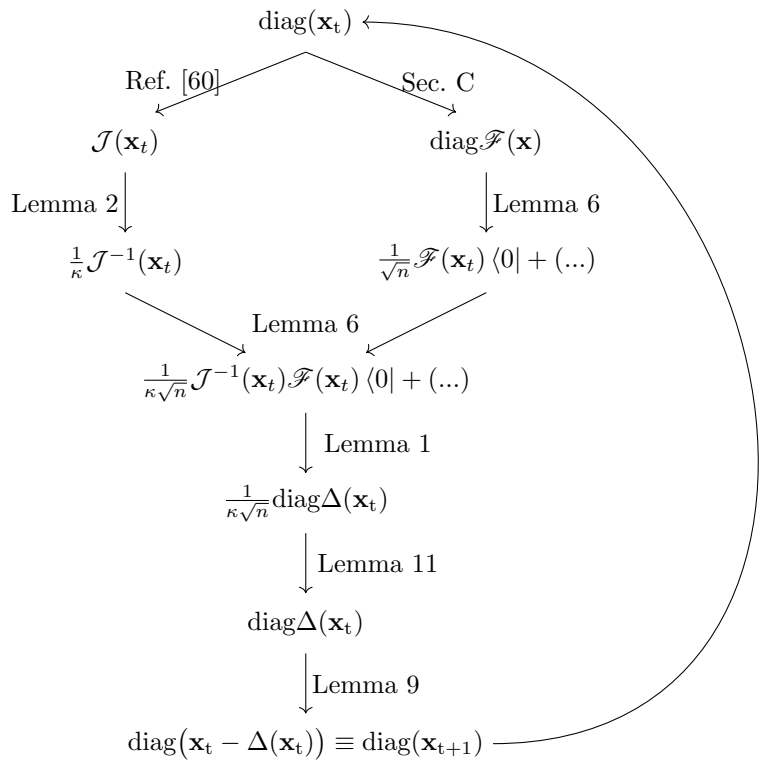
\begin{center}
    \begin{tikzpicture}[scale = 0.8]
        \node (A) at (0,5) { $\rm diag (\xbf_t)$}; 
        \draw[->] (0,4.5) -- (-2.5,3.5) ; 
        \node at (-2.2,4) {Ref.~\cite{nghiem2025quantum}}; 
        \draw[->] (0,4.5) -- (2.5,3.5);
        \node at (2.2, 4) {Sec.~\ref{sec: continuation}};
        \node at (-3,3) { $\Jcal(\xbf_t)$ }; 
        \node at (3,3) { $\rm diag  \mathscr{F}(\xbf)$   };
        \draw[->] (-3,2.5) -- (-3,1.5); 
        \node at (-4, 2) {Lemma \ref{lemma: theorem56}};
        \draw[->] (3,2.5) -- (3,1.5); 
        \node at (4, 2) {Lemma \ref{lemma: product}};
        \node at (-3,1) { $ \frac{1}{\kappa }\Jcal^{-1}(\xbf_t)$ } ;
        \node at (3,1) { $ \frac{1}{\sqrt{n}} \mathscr{F}(\xbf_t) \bra{0} + (...) $ } ;
        \draw[->] (-3,0.5) -- (-1,-0.5); 
        \draw[->] (3,0.5) -- (1,-0.5); 
        \node at (0,0) {Lemma \ref{lemma: product}};
        \node at (0,-1) {$\frac{1}{\kappa\sqrt{n}}\Jcal^{-1}(\xbf_t)  \mathscr{F}(\xbf_t) \bra{0} + (...) $ }; 
        \draw[->] (0,-1.5) -- (0,-2.5); 
        \node at (1.2,-2) {Lemma \ref{lemma: diagonal}};
        \node at (0,-3) { $\frac{1}{\kappa\sqrt{n}} \rm diag  \Delta(\xbf_t) $  }; 
        \draw[->] (0,-3.5) -- (0,-4.5); 
        \node at (1.2,-4) {Lemma \ref{lemma: amp_amp}};
        \node at (0,-5) {$\rm diag  \Delta(\xbf_t)  $};  
        \draw[->] (0,-5.5) -- (0,-6.5) ;
        \node at (1, -6) {Lemma \ref{lemma: sumencoding}};
        \node (B) at (0,-7) {$ \rm diag \big(  \xbf_t -\Delta(\xbf_t) \big) \equiv \rm diag (\xbf_{t+1}) $};
        \draw[->] (B) to[out = 0, in = 0 ,looseness = 1.5] (A);
    \end{tikzpicture}
\end{center}
Thus, beginning with some $\xbf_0$, Lemma \ref{lemma: diagonal} allows us to construct the block encoding of $ \rm diag (\xbf_0)$. Execution of the above diagram allows us to construct the block encoding of $\xbf_1$ and use this block-encoded operator to iterate the same procedure, leading to (block encoding of) $\xbf_2,...,\xbf_t,...,\xbf_T$. In order to obtain the state $\ket{\xbf_T}$, we use the block encoding of $\xbf_T$ and apply it to the state $\ket{\bf 0} \frac{1}{\sqrt{n}}\sum_{i=1}^n \ket{i-1}$. Measuring the ancilla and post-selecting on $\ket{\bf 0}$ yields $\ket{\xbf_T}$ in the remaining register, according to Definition \ref{def: blockencode}. We remark that according to Ref.~\cite{burton2009newton, pacsca2011formal}, by choosing $T = \mathcal{O}\big(\log \log \frac{1}{\epsilon}\big)$, it is guaranteed that $\xbf_T$ is $\epsilon$-close to the actual root of given nonlinear algebraic equations. We recapitulate the main result of the above procedure in the following: 
\begin{theorem}[Quantum Newton's Algorithm]
\label{thm: quantumnewton}
    Let the system of equations $\Fsr(\xbf)$ be defined as in Eqn.~\ref{nonlinearsystem}, where $f_1(\xbf),f_2(\xbf),...,f_n(\xbf)$ are one of the function types as above. Then there is a quantum algorithm that returns a quantum state $\ket{\Tilde{\xbf}}$ such that $||\ket{\Tilde{\xbf}}- \ket{\xbf}|| \leq \epsilon$, where $\ket{\xbf}$ corresponds to the solution $\xbf$ of nonlinear algebraic system $\Fsr(\xbf) = \Vec{0}$. The complexity is $\mathcal{O}\big( \Lambda^2 n \sqrt{n} \log(n)\log \frac{1}{\epsilon} \big) $
\end{theorem}
The best classical algorithm for solving nonlinear algebraic equation has scaling $\mathcal{O}(n^3)$. Thus, our quantum algorithm provides almost a quadratic speedup in the dimension $n$. Nonlinear equations capture many interesting and realistic phenomena, hence we envision that our quantum algorithm can deliver meaningful applications. Subsequently, in Section \ref{sec: application}, we shall extend our quantum Newton's algorithm in multiple directions, including solving dense linear system, probing dynamical feature of nonlinear coupled oscillators, and enhancing motion planning, which is a highly practical problem from the field of robotics.

In the following section, we proceed to describe our main quantum algorithm, describing all steps of the aforementioned diagrams in a more rigorous manner.

\section{Quantum Algorithm for Dissecting Roots}
\label{sec: quantumdissecting}
Given $n$ scalar values $x_1,x_2,...x_n \in \Rbb$ corresponding to $n$ points on the domain $\mathscr{D}$, there is a variety of methods have been proposed \cite{grover2000synthesis,grover2002creating,plesch2011quantum, schuld2018supervised, nakaji2022approximate,marin2023quantum,zoufal2019quantum} for loading classical data into amplitude of quantum state, i.e., amplitude encoding technique. In particular, the work \cite{zhang2022quantum} has provided a universal amplitude encoding strategy with the circuit depth being logarithmical in dimension. In the following, we proceed with the premise that there is an efficient circuit that generates the state, or a state that contains $(x_1,x_2,...x_n)^T $ in its first $n$ entries (could be up to some normalization factor). Such a result allows us to apply the following result from \cite{rattew2023non, guo2024nonlinear}:
\begin{lemma}[Theorem 2 in \cite{rattew2023non}]
\label{lemma: diagonal}
     Given an n-qubit quantum state specified by a state-preparation-unitary $U$, such that $\ket{\psi}_n=U\ket{0}_n=\sum^{N-1}_{k=0}\psi_k \ket{k}_n$ (with $\psi_k \in \mathbb{C}$ and $N=2^n$), we can prepare an exact block-encoding $U_A$ of the diagonal matrix $A = {\rm diag}(\psi_0, ...,\psi_{N-1})$ with $\mathcal{O}(n)$ circuit depth and a total of $\mathcal{O}(1)$ queries to a controlled-$U$ gate  with $n+3$ ancillary qubits.
\end{lemma}
to construct the block encoding of $\bigoplus_{j=1}^n x_j $. If there is a normalization factor, Lemma \ref{lemma: amp_amp} can be applied to remove it. Then from such a block encoding of $\bigoplus_{j=1}^n x_j $, we use the following central recipe from the seminar quantum singular value transformation framework: 
\begin{lemma}\label{lemma: qsvt}[\cite{gilyen2019quantum} Theorem 56 ]
\label{lemma: theorem56}  
Suppose that $U$ is an
$(\alpha, a, \epsilon)$-encoding of a Hermitian matrix $A$. (See Definition 43 of~\cite{gilyen2019quantum} for the definition.)
If $P \in \mathbb{R}[x]$ is a degree-$d$ polynomial satisfying that
\begin{itemize}
\item for all $x \in [-1,1]$: $|P(x)| \leq \frac{1}{2}$,
\end{itemize}
then, there is a quantum circuit $\tilde{U}$, which is an $(1,a+2,4d \sqrt{\frac{\epsilon}{\alpha}})$-encoding of $P(A/\alpha)$ and
consists of $d$ applications of $U$ and $U^\dagger$ gates, a single application of controlled-$U$ and $\mathcal{O}((a+1)d)$
other one- and two-qubit gates.
\end{lemma}
More specifically, we use the above lemma to transform $\bigoplus_{j=1}^n x_j \longrightarrow \bigoplus_{j=1}^n f(x_j)$. Given that a block encoding of identity matrix of any dimension is simple to prepare (see below Definition.~\ref{def: blockencode}), use Lemma \ref{lemma: sumencoding} to construct the block encoding of $\frac{1}{2} \big( \Ibb_n - \bigoplus_{j=1}^n x_j \big) = \frac{1}{2} \bigoplus_{j=1}^n \big(1-f(x_j) \big)$. As we have $|f(x)|\leq \frac{1}{2}$ for any $x\in \mathscr{D}$, this shift guarantees that the operator $\frac{1}{2} \bigoplus_{j=1}^n \big(1-f(x_j) \big) $ is positive-semidefinite in the considered domain, i.e., all the eigenvalues are greater than zero. Such a condition allows us to apply the following result from \cite{nghiem2022quantum, nghiem2024improved, nghiem2023improved}:
\begin{lemma}
\label{lemma: largestsmallest}
    Given the block encoding of a positive-semidefinite Hermitian matrix $A$ of size $n\times n$ (assumed to have $\mathcal{O}(1)$ gap between two largest eigenvalues), the largest eigenvalue can be estimated up to additive accuracy $\epsilon$ in complexity $\mathcal{O}\Big(  T_A \frac{1}{\epsilon} \big(\log n +  \log \frac{1}{\epsilon}\big)\Big)$ where $T_A$ is the complexity of producing block encoding of $A$. 
\end{lemma}
The application of the above lemma is straightforward. The block-encoded operator $\frac{1}{2}\bigoplus_{j=1}^n \big(1- f(x_j) \big)$ is positive-semidefinite, with the producing complexity as $\mathcal{O}( \deg(f) \log n)$ (due to the use of Lemma \ref{lemma: diagonal} and Lemma \ref{lemma: theorem56}). So the largest eigenvalue and smallest eigenvalue can be revealed with additive accuracy $\epsilon$ in complexity 
$$\mathcal{O}\Big(  \deg (f)  \log (n) \frac{1}{\epsilon}\big(  \log (n) + \log \frac{1}{\epsilon} \big)  \Big)$$

We remark that the spectrum of $\bigoplus_{j=1}^n  f(x_j)$ is shifted by $\frac{1}{2}$. It means that if the largest eigenvalue of $\frac{1}{2} \bigoplus_{j=1}^n \big(1-f(x_j) \big) $, is smaller than $\frac{1}{2}$, it implies that the minimum eigenvalue of $\bigoplus_{j=1}^n f(x_j) $, which is essentially 
$$\min \{ f(x_1),f(x_2),...,f(x_n) \}$$
is greater than $0$. It also indicates that the maximum eigenvalue of $ \bigoplus_{j=1}^n f(x_j)  $, which is $\max \{ f(x_1),f(x_2),...,f(x_n) \} $, is also greater than zero. In this case, there is no sign change, which implies that there is no root. On the other hand, if the largest eigenvalue of $\frac{1}{2} \bigoplus_{j=1}^n \big(1-f(x_j) \big) $, is greater than $\frac{1}{2}$, the same reasoning leads to $\min \{ f(x_1),f(x_2),...,f(x_n) \} < 0$. In this case, if $\max \{ f(x_1),f(x_2),...,f(x_n) \} \geq 0$ then there is root, otherwise if $ \max \{ f(x_1),f(x_2),...,f(x_n) \} < 0$ , then there is no root. \\

\noindent
\textbf{Discussion and generalization. } Classically, a simple solution is evaluating the sign of all $f(x_1),f(x_2),...,f(x_n)$, which takes at least $\Omega(n)$ steps. Thus, it suggests that our quantum algorithm yields a superpolynomial speedup compared to the classical counterpart. It provides a simple but quite stunning example of quantum advantage in function-related problems. We remark that the above construction considered a univariate function, and hence it is highly motivating to extend the discussion to higher dimension. Suppose without loss of generality that we are working with $M$ dimensions, and the domain of interest is $\mathscr{D} \subset [-1,1]^M$. Let $\xbf = (x_1,x_2,...,x_M) \in \mathscr{D}$ and $f(\xbf)$ refers to a multivariate function $f$ being evaluated at $\xbf$, with a further assumption that $f(\xbf)$ is multivariate polynomial, or can be approximated by some polynomial function within $\mathscr{D}$. More specifically, let $f(\xbf)$ be composed of monomials as following $f(\xbf) = \sum_{k=1}^K f_k(\xbf) = \sum_{k=1}^K a_k x_1^{k_1} x_2^{k_2} ... x_M^{k_M}$, where $k_1,k_2,...,k_n \in \Rbb$. 

We recall that in the above, we have relied on the \textit{intermediate value theorem} to determine the existence of a root. Despite there being no direct generalization of this theorem in a higher dimension, we note that the insight still holds: \\

\noindent
\textit{ If there is a change of sign of $f(\xbf)$, e.g., $\rm sign \ f(\xbf_1) \neq \rm sign \ f(\xbf_2)$ for $\xbf_1 \neq \xbf_2$, then there is at least one point where $f(\xbf) = 0$}  

The strategy turns out to be the same as we used before. We pick $n$ points $\xbf_1, \xbf_2,...,\xbf_n$ and examine if there is a change of sign among $f(\xbf_1), f(\xbf_2), ..., f(\xbf_n)$, which can be revealed from $\min \{ f(\xbf_1), f(\xbf_2), ..., f(\xbf_n) \} $ and $ \max \{ f(\xbf_1), f(\xbf_2), ..., f(\xbf_n) \} $. The only trickier point is how to construct a matrix that contains these $f(\xbf_i)$ on the diagonal, similar to the previous case. 

Let the coordinates of $\xbf_1$ be $(x_{1,1},x_{2,1},...,x_{M,1}) $. Similarly, coordinates of $\xbf_2$ is $(x_{1,2},x_{2,2},...,x_{M,2}) $, ..., of $\xbf_n$ is $(x_{1,n},x_{2,n},...,x_{M,n}) $. Recall from the above univariate case that we are provided (via amplitude encoding) with an efficient circuit that generates the state, or a state that contains $(x_1,x_2,...x_n)^T $ in its first $n$ entries. In this multivariate case, suppose via the same means, e.g., amplitude encoding, we are provided with a state containing $(x_{1,1},x_{1,2}, ..., x_{1,n})^T$, $(x_{2,1},x_{2,2}, ..., x_{2,n})^T$, ..., $(x_{M,1},x_{M,2}, ..., x_{M,n})^T$. Then Lemma \ref{lemma: diagonal} allows us to construct the block encoding of $ \bigoplus_{j=1}^n x_{i,j} $, for $i=1,2,...,M$. Then Lemma \ref{lemma: product} can be applied to obtain the transformation $ \bigoplus_{j=1}^n x_{i,j}  \longrightarrow   \bigoplus_{j=1}^nx_{i,j}^{k_i} $ for $i=1,2,...,M$. Then apply Lemma \ref{lemma: product} to construct the block encoding their products, which is $ \bigoplus_{j=1}^n x_{1,j}^{k_1} x_{2,j}^{k_2}... x_{M,j}^{k_M} $. Then we use Lemma \ref{lemma: scale} to insert the factor $a_j$, i.e., we obtain the block encoding of  $ \bigoplus_{j=1}^n  a_k x_{1,j}^{k_1} x_{2,j}^{k_2}... x_{M,j}^{k_M} $. Finally, we reppeat the above procedure to construct the same block encoding but for different $(k_1,k_2,...,k_M)$, then we use Lemma \ref{lemma: sumencoding} to construct the block encoding of $ \frac{1}{K} \bigoplus_{j=1}^n \sum_{k=1}^K a_k x_{1,j}^{k_1} x_{2,j}^{k_2}... x_{M,j}^{k_M}  = \frac{1}{K} \bigoplus_{j=1}^n f(\xbf_j)  $

The prefactor $K$ can be removed using Lemma \ref{lemma: amp_amp}. Then similar to univariate case, we can use the resultant block-encoded operator to first perform the shift $ \frac{1}{2} \Big( \Ibb_n -  \bigoplus_{j=1}^n f(\xbf_j)  \Big)$ and apply Lemma \ref{lemma: largestsmallest} to find out the largest  eigenvalue of corresponding operators, which can be used to examine whether or not there is a change of sign among $f(\xbf_1),f(\xbf_2),...,f(\xbf_n)$, which can be used to infer the existence of root of $f(\xbf)$ in the domain $\mathscr{D}$. 

For each $i=1,2,...,M$, the complexity to produce the block encoding of $ \bigoplus_{j=1}^n x_{i,j} $ is $\mathcal{O}(\log n)$. To produce the block encoding of $ \bigoplus_{j=1}^n x_{i,j}^{k_i}  $, it takes $k_i$ usage of $\bigoplus_{j=1}^n x_{i,j} $, thus the complexity is $\mathcal{O}\big( k_i \log n \big)$. The next step is producing block encoding of $\bigoplus_{j=1}^n  a_k x_{1,j}^{k_1} x_{2,j}^{k_2}... x_{M,j}^{k_M}  $, taking further $\mathcal{O}\big( (k_1+ k_2+...+k_M) \log n  \big)$, following by the construction of $\frac{1}{K} \bigoplus_{j=1}^n \sum_{k=1}^K a_k x_{1,j}^{k_1} x_{2,j}^{k_2}... x_{M,j}^{k_M} $. The complexity up to this point is $\mathcal{O}\big( K  (k_1+ k_2+...+k_M) \log n  \big)$. The application of Lemma \ref{lemma: amp_amp} to remove the factor $K$ taking further $\mathcal{O}(K)$ complexity. Finally, an application of Lemma \ref{lemma: largestsmallest} leads to the final complexity to be $\mathcal{O}\Big( K^2 (k_1+ k_2+...+k_M) \log n \big(  \log(n) +  \log \frac{1}{\epsilon}\big) \Big)$. 

We have established a quantum algorithm that can determine whether or not there is a root within a given domain $\mathscr{D}$ for a univariate $f(x)$ and multivariate function $f(\xbf)$, respectively. The above algorithms only require the knowledge of the function considered, without the oracle/black-box access to any classical data. Thus, it affirms the prospect for quantum advantage without coherent access that was put forth in \cite{nghiem2024simple} and \cite{nghiem2025quantum}. 

\section{Quantum Newton's Method for Solving Nonlinear Algebraic Equations}
\label{sec: solvingnonlinear}

As the first step of Newton's method, we need to build a linear system $\Jcal (\xbf)  \Delta = \Fsr(\xbf)$ in quantum setting. In the following, we describe a procedure for obtaining (block encoding of) $\Jcal(\xbf)$ and a operator that contains $\Fsr(\xbf)$. 

\subsection{Constructing block encoding of Jacobian $\Jcal (\xbf)$}
\label{sec: blockencodingjacobian}
The first challenge of the method above is the construction of Jacobian $\Jcal(\xbf)$, which is composed of gradient of $f_i(\xbf)$. Under the conditions as well as the type of functions mentioned earlier, the two works \cite{nghiem2024simple} have developed the following tool:
\begin{lemma}[Ref.~\cite{nghiem2024simple}]
    For $j =1,2,..,n$, let $f_i(\xbf)$ be a $n$-variable functions having the structure as defined. Suppose that there is a quantum circuit of depth $T_X$ that is the block encoding of $\rm diag(\xbf)$. Then there exists a quantum circuit of depth $\mathcal{O}\big( T_X  + \log n \big)$ that is a unitary block encoding of $\rm diag \frac{1}{M}\big( \bigtriangledown  f_j(\xbf)\big)$
    \label{lemma: diagonalgradient}
\end{lemma}
The above result combined with Lemma \ref{lemma: product} and Lemma \ref{lemma: tensorproduct} allows us to construct the block encoding of 
$$ \frac{1}{M} \ket{0}\bra{0} \otimes \rm diag \bigtriangledown f_1(\xbf), \frac{1}{M} \ket{1}\bra{1} \otimes   \rm diag \bigtriangledown  f_2(\xbf) ,..., \frac{1}{M} \ket{n-1}\bra{n-1} \otimes \rm diag \bigtriangledown f_n(\xbf) $$
Then Lemma \ref{lemma: sumencoding} can be used to construct the block encoding of $ \frac{1}{nM} \bigoplus_{j=1}^n   \rm diag \bigtriangledown f_j(\xbf)$. From here, there exists a procedure that returns the block encoding of $ \frac{1}{Mn^2} \Jcal(\xbf)$. Details can be found in Appendix \ref{sec: detailjacobian} (see Algorithm \ref{method: algorithmjacobian}). 

We remark that for any $j$, the complexity of obtaining the block encoding of projector $\ket{j-1}\bra{j-1}$  is $\mathcal{O}(\log n)$ (see Lemma \ref{lemma: projector}), thus the complexity for obtaining the block encoding of $ \frac{1}{nM }\bigoplus_{j=1}^n \rm diag \bigtriangledown f_j(\xbf)$ is $\mathcal{O}\big(nT_X + n\log n\big)$. The complexity of producing block encoding of $  \frac{1}{Mn^2} \Jcal(\xbf)$ turns out to be (asymptotically) the same $\mathcal{O} \big( nT_X + n \log n  \big)$. The factor $n$ presented here seems to be costly. In the next section, we will point out that such a scaling can be improved to $ \mathcal{O}(T_X + \log n)$ under the condition that all the functions $\{f_j(\xbf) \}_{j=1}^n$ share similar algebraic form.

\subsection{Constructing block encoding of $\rm diag \Fsr(\xbf)$}
\label{sec: blockencodingF}
This section aims at building the block encoding of $\rm diag \Fsr(\xbf)$, with a reminder that $\Fsr(\xbf)  = \big(f_1(\xbf), f_2(\xbf), ..., f_n(\xbf) \big)^T$ and thus $ \rm diag \Fsr(\xbf) = \bigoplus_{j=1}^n f_j(\xbf)$. Again, the detail is quite technical and lengthy, we leave it in the Appendix \ref{sec: continuation}, and we only provide a few essential results in the following. For each of functions provided in Eqn.~\ref{eqn: function}, and for each $i$ (=1,2,..,K), suppose that we are provided with a unitary $U_{a_i}$ that generates the state $\ket{a_i}=  (\sqrt{a_{i,1}}, \sqrt{a_{i,2}},..., \sqrt{a_{i,n}})^T$ (assuming to have unit norm without loss of generalization). Recall further from Lemma \ref{lemma: diagonalgradient} that we have the block encoding of $\rm diag (\xbf)$ with complexity $T_X$. Given these abilities, it is possible to construct a block encoding of $\ket{0}\bra{0} f_1(\xbf), \ket{1}\bra{1}f_2(\xbf) ,..., \ket{n-1}\bra{n-1} f_n(\xbf) $ in complexity $\mathcal{O}\Big( T_X + \log n \Big)$. Then Lemma \ref{lemma: sumencoding} allows us to construct the block encoding of $\frac{1}{n}\Big(\ket{0}\bra{0} f_1(\xbf) + \ket{1}\bra{1}f_2(\xbf) + ... +  \ket{n-1}\bra{n-1} f_n(\xbf)   \Big) = \frac{1}{n}\bigoplus_{j=1}^n f_j(\xbf) $. The factor $n$ can be removed using Lemma \ref{lemma: amp_amp}. Assume that $K =\mathcal{O}(1)$, the total complexity for producing $\bigoplus_{j=1}^n f_j(\xbf)$ is $\mathcal{O}\Big(  n^2 (T_X + \log n\Big)$, which is apparently very costly due to the scaling $n^2$. It turns out that such a scaling can be significantly improved if for all $j=1,2,...,n$, $f_j(\xbf)$ shares similar form, for instance: 
\begin{align}
    \Fsr(\xbf) = \begin{cases}
        f_1(\xbf) \\
        f_2(\xbf) \\
        \vdots \\
        f_n(\xbf)
    \end{cases} = \begin{cases}
        \sum_{i=1}^K a_{i,1}^1 x_1^i + a_{i,2}^1 x_2^i + ... + a_{i,n}^1 x_n^i \\
         \sum_{i=1}^K a_{i,1}^2 x_1^i + a_{i,2}^2 x_2^i + ... + a_{i,n}^2 x_n^i \\
          \vdots \\
        \sum_{i=1}^K a_{i,1}^n x_1^i + a_{i,2}^n x_2^i + ... + a_{i,n}^n x_n^i \\
    \end{cases}
\end{align}
Additionally, for all $i$, if we are provided with a unitary $U_{A_i}$ that generates a state 
$$\sim \big( \sqrt{a^1_{i,1}}, \sqrt{a^1_{i,2}}, ..., \sqrt{a^1_{i,n}}, \sqrt{a^2_{i,1}}, \sqrt{a^2_{i,2}}, ..., \sqrt{a^2_{i,n}}, ..., \sqrt{a^n_{i,1}}, \sqrt{a^n_{i,2}}, ..., \sqrt{a^n_{i,n}} \big)^T  $$
(again we remind that $\sim$ refers to a possible normalization/multiplication factor). We remark that, in principle, as long as all the entries in the above state are known, then it can be efficiently constructed, e.g., by a circuit of depth $\mathcal{O}(\log n^2) = \mathcal{O}( \log n)$ by the method in \cite{zhang2022quantum}. Then there is a procedure that returns a block encoding of $\bigoplus_{j=1}^nf_j(\xbf) = \rm diag \Fsr(\xbf)$ in complexity $\mathcal{O}\big( T_X + \log n\big)$. 

Quite surprisingly, the above improvement suggests a potential improvement to the previous section. Recall that in the previous section (the first two steps of Algorithm \ref{method: algorithmjacobian} in Appendix \ref{sec: detailjacobian}), we construct the block encoding of $ \frac{1}{Mn}\bigoplus_{j=1}^n \rm diag \big(  \bigtriangledown f_j(\xbf) \big) $, where we utilized the block encoding of $ \ket{j-1}\bra{j-1}\otimes \rm diag \big(  \bigtriangledown f_j(\xbf) \big)$ for all $j=1,2,..,n$ with Lemma \ref{lemma: sumencoding}, thus leading to $\mathcal{O}(n)$ complexity scaling (see below Step (7) of \ref{method: algorithmjacobian}). In the Appendix \ref{sec: continuation}, we extend our technique and directly show that if for all  $j=1,2,...,n$, and  $\{ f_j(\xbf) \}$ shares similar form, then the block encoding of operator $\frac{1}{M} \bigoplus_{j=1}^n \rm diag\big( \bigtriangledown  f_j(\xbf) \big)$ can be constructed in complexity $\mathcal{O}\Big(  T_X + \log n \Big)$. Thus, the complexity to produce the block encoding of  $ \frac{1}{Mn } \Jcal(\xbf)$ can be improved to $\mathcal{O}\big( T_X + \log n \big)$. In the following, we proceed with the scenario where all $\{ f_j(\xbf)\}_{j=1}^n$ shares the same form, so we can take advantage of the improved complexity.

\subsection{ Constructing $\Delta(\xbf) = \Jcal^{-1}(\xbf) \Fsr(\xbf)$ }
\label{sec: blockencodingdelta}
Recall that the previous two sections yield the block encoding of $\frac{1}{n^2}\Jcal(\xbf)$ and of $\rm diag \Fsr(\xbf)$ from the block encoding of $\rm diag (\xbf)$, which is assumed to have for now. We first use Lemma \ref{lemma: product} to construct the block encoding of $ \rm diag \Fsr(\xbf) \cdot H^{\otimes \log n} $. Given that $  \rm diag \Fsr(\xbf) = \bigoplus_{j=1}^ n f_j(\xbf)$, so we obtain the block encoding of    $ \sum_{k=1}^n \frac{1}{\sqrt{n}} f_j(\xbf) \ket{j-1} \bra{0} + (...) = \frac{1}{\sqrt{n}} \Fsr(\xbf) \bra{0} + (...)$ where $(...)$ denotes irrelevant parts.  

From the block encoding of $ \frac{1}{M n} \Jcal(\xbf)$, we can leverage the matrix inversion technique \cite{gilyen2019quantum, childs2017quantum} to obtain the block encoding of $\frac{1}{\kappa} \Jcal^{-1} (\xbf)$ where $\kappa$ is the conditional number of $\Jcal(\xbf)$ -- which can be revealed by using the tool provided earlier, Lemma \ref{lemma: largestsmallest}, to find the largest and smallest singular value of $ \frac{1}{M n} \Jcal(\xbf) $. More precisely, we invert the matrix $ \frac{1}{M n} \Jcal(\xbf) $, and obtain the block encoding of $\Big( \frac{1}{M n} \Jcal(\xbf)\Big)^{-1} =  M n \frac{1}{\kappa'} \Jcal(\xbf)^{-1} $ where $\kappa'$ is the conditional number of $  \frac{1}{M n} \Jcal(\xbf)$. As it only differs from $\Jcal(\xbf)$ by a factor, we have $\kappa' = \kappa Mn$, thus the factor $Mn$ vanishes. 

As a next step, we use Lemma \ref{lemma: product} to construct the block encoding of:
\begin{align}
    \frac{1}{\kappa} \Jcal^{-1} (\xbf) \cdot  \frac{1}{ \sqrt{n}}\Big( \Fsr(\xbf) \bra{0} + (...) \Big) =   \frac{1}{\kappa \sqrt{n}} \Jcal^{-1} (\xbf)  \Fsr(\xbf) \bra{0}  + (...)
\end{align}
If we define the linear system as $\Jcal (\xbf) \Delta(\xbf) = \Fsr(\xbf)$, then the solution is $\Delta(\xbf) = \Jcal^{-1}(\xbf) \Fsr(\xbf)$. Then the above block-encoded operator is $\frac{1}{\sqrt{n}\kappa} \Delta(\xbf) \bra{0} + (...)$ which is essentially an $n \times n$ matrix containing $\frac{1}{\sqrt{n}\kappa}\Delta(\xbf)$ as the first columns. Lemma \ref{lemma: diagonal} allows us to use such block encoding to construct the block encoding of $\rm diag \frac{1}{\sqrt{n}\kappa } \Delta(\xbf)$. With the knowledge of $\kappa$, we can use Lemma \ref{lemma: amp_amp} to remove the factor $\sqrt{n}\kappa$, resulting in the block encoding of $\rm diag \Delta(\xbf)$. 

Recall that the complexity to obtain the block encoding of $\frac{1}{Mn}\Jcal(\xbf)$ is $\mathcal{O}( T_X +  \log n)$. To obtain a block encoding of its inverse, $\frac{M n^2}{\kappa} \Jcal^{-1} (\xbf)  $,  we need to use Lemma \ref{lemma: theorem56} with an appropriate polynomial that approximates the inverse of singular values of $\frac{1}{Mn}\Jcal(\xbf) $. According to \cite{gilyen2019quantum} (Corollary 69), this polynomial has a degree $\mathcal{O}\big( Mn \kappa \log \frac{1}{\epsilon}  \big)$. So the complexity of producing the $\epsilon$-approximated block encoding of $\frac{1}{\kappa} \Jcal^{-1} (\xbf)  $ is $\mathcal{O}\Big( \kappa M n \log \frac{1}{\epsilon}   \big( nT_X + n\log n \big)   \Big)$. 

In addition, the complexity of producing the block encoding of $  \rm diag \big( \Fsr(\xbf) \big)$ is $\mathcal{O}\big( T_X + \log n\big)$. Therefore, the complexity of producing the block encoding of $\frac{1}{\sqrt{n}}  \Fsr(\xbf)\bra{0} + (...) $ is also $\mathcal{O}(n T_X +n \log n)$. So, the total complexity of producing the block encoding of $  \frac{1}{\kappa \sqrt{n}}\Delta(\xbf) \bra{0} + (...) $ is $ \mathcal{O}\Big( \kappa M n \log \frac{1}{\epsilon}  \big( T_X + \log n\big)  \Big)  $. The final step is using Lemma \ref{lemma: diagonal} to transform the block encoding of $\frac{1}{\sqrt{n}\kappa} \Delta(\xbf) \bra{0} + (...)$ into $\frac{1}{\sqrt{n}\kappa    } \rm diag \Delta(\xbf)$ and then Lemma \ref{lemma: amp_amp} to multiply the block-encoded operator $ \frac{1}{\kappa \sqrt{n}}\Delta(\xbf) \bra{0} + (...)  $ with $\kappa \sqrt{n}$, resulting in complexity $\mathcal{O}\Big(  \kappa^2  M n\sqrt{n} \log \frac{1}{\epsilon}  \big( T_X + \log n\big)  \Big) $.

\subsection{Quantum Newton's method}
\label{sec: quantumnewtonmethod}
This section completes the construction of quantum Newton method that returns a quantum state $\ket{\xbf}$ corresponding to the solution $\xbf$ of the nonlinear algebraic equations defined earlier. A formal description of quantum Newton's method is the following.
\begin{method}[Quantum Newton's Method]
    \label{method: quantumnewtonmethod}
\end{method}
\begin{enumerate}
    \item Use a random unitary $U_0$ that generates $\ket{\xbf_0} \equiv \xbf_0$, i.e., $U_0 \ket{0} = \xbf_0$ where we abuse $\ket{0}$ denotes specifically the first computational basis state of a $n$-dimensional Hilbert space. Note that the first column of $U_0$ is $\xbf_0$, i.e., $U_0 = x_0 \bra{0} + (...)$ where $(...)$ denotes the redundant part. 
    \item Use Lemma \ref{lemma: diagonal} with $U_0$ to construct the block encoding of $\rm diag (\xbf_0)$ 
    \item From the block encoding of $\rm diag (\xbf_0)$, use the method in the previous section to construct the block encoding of $\rm diag \Delta(\xbf_0)$

    \item Use Lemma \ref{lemma: sumencoding} to construct the block encoding of $\frac{1}{2} \Big( \rm diag \xbf_0 - \rm diag \Delta(\xbf_0) \Big)$ 
     \item Use Lemma \ref{lemma: amp_amp} to multiply the above block-encoded operator with $2$. Note that $\xbf_0 - \Delta(\xbf_0) = \xbf_1$, according to Newton's method. 
    \item Repeat from Step (2) with $\rm diag \xbf_1$ in place of $\rm diag \xbf_0$. Iterate the procedure for a total of $T$ times, we obtain a unitary block encoding $\rm diag (\xbf_T)$.
    \item Apply such a unitary to $\ket{\bf 0} \frac{1}{\sqrt{n}} \sum_{j=1}^n \ket{j-1}$, according to Definition \ref{def: blockencode},  we obtain $ \frac{1}{\sqrt{n}}\ket{\bf 0}  \xbf_T + \ket{\rm Garbage} $. Measure the first register and post-select on $\ket{\bf 0}$, we obtain the state $\ket{\xbf_T} =  \frac{\xbf_T}{||\xbf_T||}$.
\end{enumerate}

\subsection{ Discussion and extension}
\label{sec: disscusionandgeneralizationnewtonmethod}

\noindent
\textbf{Complexity.} First we examine the complexity of our quantum Newton's method. Since the method is iterative, we can use induction strategy. At $t$-th time step, suppose that we begin with a unitary $U_t$ that contains $\xbf_t$ in its first column, and let $T_{X_t}$ denotes the complexity of $U_t$. Recall previously that the complexity to produce the block encoding of $\sum_{p=1}^n \Delta(\xbf_{t+1}) \bra{p-1}   $ is $\mathcal{O}\Big( \kappa^2(\xbf_t)n\sqrt{n} \log \frac{1}{\epsilon}   \big( T_{X_t} + \log n \big)   \Big) $, where $\kappa(\xbf_t)$ is the conditional number of $\frac{1}{Mn^2}\Jcal(\xbf_t)$. The first two steps of Algorithm \ref{method: quantumnewtonmethod} has complexity $\mathcal{O}\big(  T_{X_t} +  \log n)$. Step (4) and (5) uses two block encoding of $ \sum_{p=1}^n \Delta(\xbf_t)\bra{p-1} $ and of $U_t$, therefore, the total complexity for obtaining the unitary $U_{t+1}$ that contains $\xbf_{t+1}$ in its first column is $\mathcal{O}\Big( \kappa^2(\xbf_t)n\sqrt{n}\log \frac{1}{\epsilon}   \big( n T_{X_t}+ \log n \big)   \Big) $. We recall that at the beginning, there was an assumption on the lower bound $\frac{1}{\Lambda}$ of the smallest singular value, in magnitude, of $\Jcal(\xbf)$ for all $\xbf \in \mathscr{D}$. Thus, it implies that the conditional number $\kappa(\xbf_t) \leq \Lambda$. So the complexity for obtaining $U_{t+1}$ is $T_{X_{t+1} } =  \mathcal{O}\Big( \Lambda^2 n\sqrt{n} \log \frac{1}{\epsilon}   \big(  T_{X_t}+ \log n \big)  \Big) $. Similarly, we have $ T_{X_t} = \mathcal{O}\Big( \Lambda^2 n\sqrt{n} \log \frac{1}{\epsilon}   \big(  T_{X_{t-1}}+ \log n \big)  \Big)  $. So $T_{X_{t+1} } =  \mathcal{O}\Big( \big( \Lambda^2 n\sqrt{n} \log \frac{1}{\epsilon} \big)^2 T_{X_{t-1}}+  \big(\Lambda^2 n \sqrt{n} \log \frac{1}{\epsilon} \big)^2  \log n  + \big( \Lambda^2 n \sqrt{n} \log \frac{1}{\epsilon } \big) \log(n)\Big)$. Again replacing $T_{X_{t-1}} = \mathcal{O}\Big( \Lambda^2\ n\sqrt{n} \log \frac{1}{\epsilon}  T_{X_{t-2}} + 
 \Lambda^2 n \sqrt{n} \log(n) \log \frac{1}{\epsilon} \Big) $, so we have $T_{X_{t+1} } = \mathcal{O}\Big(  \big( \Lambda^2 n\sqrt{n} \log \frac{1}{\epsilon} \big)^3 T_{X_{t-2}} + \big( \Lambda^2 n \sqrt{n}\log\frac{1}{\epsilon}\big)^3   \log(n)   + \big(\Lambda^2 n\sqrt{n} \log \frac{1}{\epsilon} \big)^2  \log n  +  \Lambda^2 n\sqrt{n} \log(n) \log \frac{1}{\epsilon }     \Big)  $. Continuing until $t=0$, we have the total complexity is: $ T_{X_{t+1}} = \mathcal{O}\Big( \big( \Lambda^2 n\sqrt{n} \log \frac{1}{\epsilon} \big)^{t+1} + \log(n)  \cdot \big( \sum_{k=0}^{t+1} (\Lambda^2 n \sqrt{n}\log \frac{1}{\epsilon} )^k  \big)  \Big) = \mathcal{O}\Big( \big(  \Lambda^2 n\sqrt{n} \log \frac{1}{\epsilon} \big)^{t+1} \log(n) \Big)$. 
 
 Thus, for a total of $T$ iterations, the complexity is $\mathcal{O}\Big( \big(  \Lambda^2 n\sqrt{n} \log \frac{1}{\epsilon} \big)^{T} \log n  \Big) $.  The exponential dependence on $T$ seems inefficient at first. However, a few works have investigated the performance of Newton's method \cite{burton2009newton, pacsca2011formal}, and it turns out that in order to for $\xbf_T$ to be $\epsilon$-close to the real root, $T = \log \log \frac{1}{\epsilon}$ suffices. So the complexity of $U_T$ is $\mathcal{O}\Big(   \Lambda^2 n\sqrt{n} \log^2 \frac{1}{\epsilon}  \log(n)  \Big) $. At the final step (7) of Algorithm \ref{method: quantumnewtonmethod}, we need to make measurement and post-select on $\ket{\bf 0}$. The success probability is $\frac{||\xbf_T||^2}{n} = \mathcal{O}(1)$, where we have used the fact that each entries of $\xbf_T$ is of $\mathcal{O}(1)$. Hence the total complexity of obtaining $\ket{\xbf_T}$ is $\mathcal{O}\Big(   \Lambda^2 n\sqrt{n} \log^2 \frac{1}{\epsilon}   \log(n)  \Big)  $. 

\noindent
\textbf{Comparing to classical Newton's method. } At $t$-th time step, the best classical approach needs to construct the Jacobian $\Jcal(\xbf_t)$ and then perform the matrix inversion to solve $\Jcal(\xbf_t) \Delta_t = \Fsr(\xbf_t)$. As far as our understanding, the best classical matrix inversion method for a dense matrix has scaling $\mathcal{O}(n^3)$, thus the total complexity for classical Newton's method is $\mathcal{O}\big( T n^3\big)$. Comparing the two complexity, we can see that there is almost a quadratic speed-up in the dimension $n$. 

\noindent
\textbf{Extension to Levenberg-Marquardt algorithm.} As we mentioned earlier, the Newton's method shares many recipes with the Levenberg-Marquardt algorithm, and thus our quantum Newton's method outlined above can be extended to quantum Levenberg-Marquardt in a straightforward manner. We recall that in the Levenberg-Marquardt algorithm, at $t$-th time step, the linear system is defined as $ \big(  \Jcal(\xbf_t)^T \Jcal(\xbf_t) + \lambda \Ibb \big) \Delta_t = -\Jcal^T (\xbf_t) \Fsr(\xbf_t)$. Section \ref{sec: blockencodingjacobian} outputs the block encoding of $\sim \Jcal(\xbf_t)$ (the symbol $\sim$ denotes some multiplication factor, more precisely). Given that the identity matrix $\Ibb$ of any dimension can be simply block-encoded, we can use Lemma \ref{lemma: scale} to construct the block encoding of $\lambda \Ibb$, then use Lemma \ref{lemma: sumencoding} to construct the block encoding of $\sim   \Jcal(\xbf_t)^T \Jcal(\xbf_t) + \lambda \Ibb$. We can use the procedure of Section \ref{sec: blockencodingdelta}, but instead of inverting $\Jcal(\xbf_t)$, we simply construct the block encoding of $\sum_{p=1}^n -\Jcal^T(\xbf_t) \Fsr(\xbf_t) \bra{p-1}$. Then again we invert $\Jcal(\xbf_t)^T \Jcal(\xbf_t) + \lambda \Ibb $, resulting in the block encoding of $\sim \big( \Jcal(\xbf_t)^T \Jcal(\xbf_t) + \lambda \Ibb\big)^{-1} \sum_{p=1}^n -\Jcal^T(\xbf_t) \Fsr(\xbf_t) \bra{p-1} = \sum_{p=1}^n \Delta_t \bra{p-1}$. Then we use such result, combined with Lemma \ref{lemma: sumencoding}, and the unitary block encoding $U_t$ of $\xbf_t \bra{0} + (...)$ to construct the block encoding of $\sim \big( \xbf_t + \Delta_t\big) \bra{0} + (...)$, and iterate the procedure $T$ times. The complexity of this quantum Levenberg-Marquardt algorithm can be shown to be asymptotically similar to quantum Newton's method, as most recipes are the same.

\section{Application and Implication}
\label{sec: application}
Our quantum Newton's method yields a unitary block encoding of $\rm diag (\xbf)$ -- which can be used to obtain the quantum state $\ket{\xbf} = \frac{\xbf}{||\xbf||}$. In the following we discuss from a broader perspective, pointing out several directions in which we can apply our quantum Newton method.  \\

\noindent
\textbf{Solving linear system.} We recall that in Figures \ref{fig: fig1} and \ref{fig: fig2}, it was illustrated that solving nonlinear algebraic systems is equivalent to finding the intersection of curves, surfaces, or more generally, algebraic varieties. Mathematically, a linear algebraic system can be viewed as a special case of nonlinear algebraic systems in which higher-order terms vanish. The problem of solving linear systems is then equivalent to finding the intersection of planes, which is among the most basic algebraic varieties.
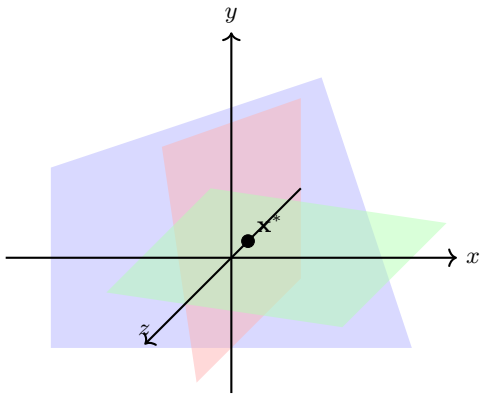
\begin{figure}[H]
    \centering
    \begin{tikzpicture}[scale=1.2]

    \begin{scope}[canvas is xy plane at z=0]
        \fill[blue!30,opacity=0.5] (-2,-1) -- (2,-1) -- (1,2) -- (-2,1) -- cycle;
    \end{scope}

    \begin{scope}[canvas is yz plane at x=0]
        \fill[red!30,opacity=0.5] (-1,-2) -- (1,-2) -- (2,2) -- (-1,1) -- cycle;
    \end{scope}

    \begin{scope}[canvas is xz plane at y=0]
        \fill[green!30,opacity=0.5] (-1,-2) -- (2,-1) -- (2,2) -- (-1,1) -- cycle;
    \end{scope}

    \draw[thick,->] (-2.5,0,0) -- (2.5,0,0) node[right] {\small $x$};
    \draw[thick,->] (0,-1.5,0) -- (0,2.5,0) node[above] {\small $y$};
    \draw[thick,->] (0,0,-2) -- (0,0,2.5) node[above] {\small $z$};

    \filldraw[black] (0.3,0.3,0.3) circle (2pt) node[above right] {\small $\mathbf{x}^*$};

\end{tikzpicture}
    \caption{ \textit{An 3-d illustration of solving linear systems. Three planes with color purple, red, green features three algebraic variety $f_1(\xbf) = 0$, $f_2(\xbf) = 0$, $f_3(\xbf) = 0$. Solution to corresponding linear systems is the intersection of these three planes (black dot) } }
    \label{fig: linearequation}
\end{figure}

Consequently, our quantum Newton's algorithm can be applied to solve linear systems with a complexity of $\mathcal{O}\Big( \kappa^2 n \sqrt{n }\log \frac{1}{\epsilon} \Big)$. Notably, our method is capable of handling dense linear systems, where the best classical algorithm has a complexity of $\mathcal{O}(n^3)$. This means our quantum linear solver achieves an almost quadratic speedup in the dimension $n$. We remark that various quantum linear solver proposals exist. However, most of these approaches either assume black-box or oracle access to the given matrix \cite{harrow2009quantum, childs2017quantum, wossnig2018quantum, nghiem2025new}, rely on heuristic methods \cite{huang2019near}, or require the matrix to have a special structure \cite{zhang2022quantum}. Recently, \cite{nghiem2025quantum} introduced another quantum linear solver, based on gradient descent, that does not require coherent access. In the dense case, their complexity is $\mathcal{O}(n^2)$, which means that our approach is faster by a factor of $\sqrt{n}$. Thus, our work contributes to the growing body of research \cite{zhang2022quantum} \cite{nghiem2025quantum}  that demonstrates the advantages of quantum linear solvers without the need of coherent quantum access to classical information. \\

\noindent
\textbf{Determining equilibrium state of 1-d nonlinear coupled masses} This is a popular physical model in 1-d, with generalization to higher dimension being straightforward by adding extra coordinates. In this setting, a system of masses $\mathscr{M}_1,\mathscr{M}_2,...,\mathscr{M}_n$ is connected by springs. If two masses at two ends are connected to the wall, the system is called open, with a total of $n+1$ springs. On the other hand, if they are connected to each other, the system is then called closed, and there are $n$ springs (see the figure below showing a closed system with 6 masses)\\
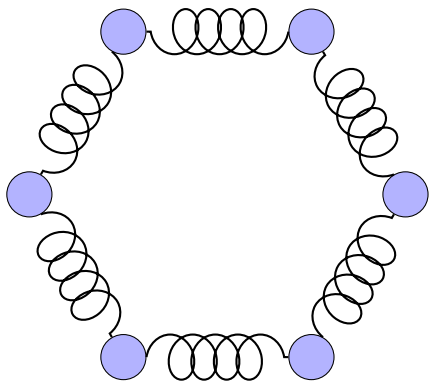
\begin{center}
\begin{tikzpicture}

    \foreach \i in {0, 1, 2, 3, 4, 5} {
        \node[circle, draw, fill=blue!30, minimum size=6mm] (m\i) at ({\i*60}:2.5cm) {};
    }

    \foreach \i in {0, 1, 2, 3, 4, 5} {
        \pgfmathtruncatemacro{\next}{mod(\i+1,6)}
        \draw[thick,decorate,decoration={coil,aspect=0.8,segment length=3mm,amplitude=3mm}] (m\i) -- (m\next);
    }
\end{tikzpicture}
\end{center}
In the following, we consider the closed boundary case for simplicity, however, extension to open boundary is straightforward. Let the spring constants be $k_1,k_2,...,k_{n}$. In the simplest setting, the force $F$ enacted by the spring obeys Hooke's law $F = k \ \Delta x$ where $\Delta x$ is the displacement -- thus the force is linear in the displacement, exhibiting the simplest model. A more realistic model accounts for many other factors, for example, the material structure of the spring, in which case the force further includes non-linear terms, e.g., $F =\sum_{j=1}^K k_j (\Delta x)^j $. 

Let $x_1,x_2,...,x_n$ denote the displacement of $m_1,m_2,...,m_n$ from the position where all springs are unstretched. In equilibrium, Newton's second law applies, which means that for each mass $m_i$, the total force on it is zero. We thus have the following nonlinear algebraic systems:
\begin{align}
  \Fsr(\xbf) = \begin{cases}
        f_1(\xbf) \\
        f_2(\xbf) \\
        \vdots \\
        f_n(\xbf)
    \end{cases} = \begin{cases}
    k_1 (x_2 - x_1 ) +  ... + k_K (x_2-x_1)^K - k_1 (x_1-x_n) - ... - k_K (x_1-x_n)^K = 0, \\
    k_1 (x_3 - x_2 ) + ... + k_k (x_3-x_2)^K - k_1 (x_2 - x_1 )  - ... -k_K (x_2-x_1)^K= 0, \\
    \vdots \\
   k_1 (x_1-x_n) + ... + k_K (x_1-x_n)^K - k_1 (x_n - x_{n-1})  - ... - k_K (x_n-x_{n-1})^K  = 0.
\end{cases}
\end{align}
Observe that each $f_j(\xbf)$ from the above falls exactly into the types of function we considered in Eqn.~\ref{eqn: function}, thus, our quantum Newton's method can be applied. The outcome of our algorithm is either a block encoding of $\rm diag (\xbf)$, or a quantum state $\ket{\xbf}$ that corresponds to the equilibrium of the system. To determine the potential energy of the system at equilibrium, we note that 
\begin{align}
    V_{\rm equilibrium} = \sum_{i=1}^{n} \sum_{p=1}^K  \frac{1}{(p+1)} k_{i,j} (x_{i+1}-x_i)^{p+1} 
\end{align}
where in the above summation, $x_{n+1} = x_1$ due to the closed boundary. For some $p$, we consider the term $\sum_{i=1}^n \frac{1}{p+1} k_{i,p} (x_{i+1}- x_i)^{p+1} $. Denote $\xbf = (x_1,x_2,...,x_n)^T$, then we have that:
\begin{align}
\label{eqn: circulant}
    \begin{pmatrix}
        x_2-x_1 \\
        x_3-x_2 \\
        \vdots \\
        x_1-x_n 
    \end{pmatrix} =\begin{pmatrix}
        -1 & 1 & 0 & \cdots & 0 \\
        0 & -1 & 1 & \cdots &0 \\
        \vdots & \vdots & \vdots & \ddots & \vdots \\
        1 & 0 & 0 & \cdots & -1
    \end{pmatrix} \begin{pmatrix}
        x_1 \\
        x_2 \\
        \vdots \\
        x_n
    \end{pmatrix} \equiv  C \xbf 
\end{align}
where $C$ is a $n\times n$ circulant matrix. In the Appendix \ref{sec: circulant}, we show that such a circulant matrix can be block-encoded with a circuit of depth $\mathcal{O}(\log n)$. To apply such a result to find the energy, we take the block encoding of $\rm diag (\xbf)$, which is $\bigoplus_{i=1}^n x_i $ , and use Lemma \ref{lemma: product} to construct the block encoding of $\rm diag (\xbf) \cdot (H^{\otimes \log n}) = \frac{1}{\sqrt{n}} \sum_{i=1}^n x_i \ket{i-1}\bra{0} + (...) $ where $(...)$ denotes the irrelevant part. Now we use Lemma \ref{lemma: product} with the block encoding of $C$ again to construct the block encoding of $C \Big( \frac{1}{\sqrt{n}} \sum_{i=1}^n x_i \ket{i-1}\bra{0} + (...) \Big) $, and note that $\sum_{i=1}^n x_i \ket{i-1} = \xbf  $ , so we obtain the unitary block encoding of $\frac{1}{\sqrt{n}} C\xbf \bra0  + (...) $. With $C\xbf$ be as above, this unitary block encoding can be used within the Lemma \ref{lemma: diagonal} to construct the block encoding of $\rm diag \frac{1}{\sqrt{n}}\Big( x_2-x_1, x_3-x_2, ..., x_1 -x_n \Big)^T$.  Then we use $p$ copies of such a block encoding with Lemma \ref{lemma: product} to construct the block encoding of $\rm diag \frac{1}{\sqrt{n^{p+1}}}\Big( (x_2-x_1)^{p+1}, (x_3-x_2)^{p+1}, ..., (x_1 -x_n)^{p+1} \Big)^T $. From such the block encoding, we repeat the same as step (7) of Algorithm \ref{method: quantumnewtonmethod}, as we use it to apply to the state $\ket{\bf 0} \frac{1}{\sqrt{n}} \sum_{i=1}^n \ket{i-1}$. The resultant state is then 
$$\ket{\Phi_1}= \ket{\bf 0}\frac{1}{\sqrt{n^{p+1}}} \sum_{i=1}^n (x_{i+1}-x_i)^{p+1}\ket{i-1}  + \ket{\rm Garbage}$$ 
By preparing another state $ \ket{\Phi_2} = \ket{\bf 0}\frac{1}{\sqrt{n}} \sum_{i=1}^n \ket{i-1}$, which is simple with unitary $\Ibb \otimes H^{\otimes \log n}$ , we can use Hadamard test or SWAP test algorithm to estimate $\braket{\Phi_2,\Phi_1} = \frac{1 }{\sqrt{n^{p+2}}} \sum_{i=1}^n (x_{i+1}-x_i)^{p+1} $, where we have used the property that $\ket{\rm Garbage}$ is orthogonal to $\ket{\bf 0}\frac{1}{\sqrt{n}} \sum_{i=1}^n \ket{i-1} $ (see Def.~\ref{def: blockencode} and Eqn.~\ref{eqn: action}). Repeating the same procedure to estimate $\frac{1 }{\sqrt{n^{p+2}}} \sum_{i=1}^n (x_{i+1}-x_i)^{p+1} $ for $p=1,2,...,K$, then we can estimate the value of $V_{\rm equilibrium}$. \\

\noindent
\textbf{Probing the dynamics of nonlinear coupled masses. } The above result motivates us to go beyond the equilibrium regime and investigate the dynamics of such a nonlinear coupled masses system, which is governed by Newton's law of motion as following: 
\begin{align}
\label{eqn: newtonequation}
    \begin{cases}
    k_1 (x_2 - x_1 ) +  ... + k_K (x_2-x_1)^K - k_1 (x_1-x_n) - ... - k_K (x_1-x_n)^K = \mathscr{M}_1 x_1'' \\
    k_1 (x_3 - x_2 ) + ... + k_k (x_3-x_2)^K - k_1 (x_2 - x_1 )  - ... -k_K (x_2-x_1)^K= \mathscr{M}_2 x_2'' \\
    \vdots \\
   k_1 (x_1-x_n) + ... + k_K (x_1-x_n)^K - k_1 (x_n - x_{n-1})  - ... - k_K (x_n-x_{n-1})^K  = \mathscr{M}_n x_n''
\end{cases}
\end{align}
subjected to some initial condition. For $K=1$, the system reduces to linear regime. An efficient quantum simulation algorithm for the dynamics of such a system was outlined in \cite{babbush2023exponential}. Their main idea is to map the dynamical equations into Schrodinger's equation, for which they can leverage many advances in quantum simulation \cite{low2017optimal,low2019hamiltonian,berry2007efficient,berry2012black,berry2014high,berry2015hamiltonian}, given oracle access to necessary information. For $K>1$, the model becomes nonlinear, and it is unclear to us whether there is any correspondence between the dynamics of this system with any system that can be simulated. In the following, we take a different route, based on the classical ordinary differential equations solver. 

Suppose that we are interested in the simulation up to time $\mathscr{T}$. We break the time interval $[0,T]$ into subintervals $[0,\Delta, 2\Delta, ..., N\Delta \equiv \mathscr{T}]$. According to Lemma 1 of \cite{childs2021high}, the following so-called central finite difference formula provides an approximation of second-order derivative as $f''(t) \approx \frac{1}{\Delta^2} \sum_{j=-M}^M r_j f(t + j \Delta)$ with the coefficients:
\begin{align}
    r_j = \begin{cases}
        \frac{2(-1)^{j+1} (M!)^2}{ j^2 (M-j)! (M+j)!}  \ \text{for } \  j \in [1,2,...,M] \\
        - 2 \sum_{j=1}^M r_j \ \text{for} \ j= 0\\
        r_{-j} \ \text{for} \ j \in -[1,2,...,M]
    \end{cases}
\end{align}
The error approximation for the above choice is roughly as $\mathcal{O}\big( \Delta^{2M}\big)$. Using the above formula, at a given time step $m \Delta  $ (for $m=0,1,2,...,N$), for all $i=1,2,...,n$, we approximate $x_i'' (m\Delta) \approx \frac{1}{\Delta^2} \sum_{j=-M}^M r_j x_i \big( (m+j) \Delta \big)  $. For note that for $m -j < 0$, we simply set $r_j = 0$. So, at $m\Delta$ time step, we arrive at a system:
\begin{align}
    \begin{cases}
    k_1 \big( x_2(m\Delta ) - x_1 (m\Delta) \big) +  ... + k_K (x_2(m\Delta) -x_1(m\Delta))^K - k_1 (x_1-x_n) - ... - k_K (x_1-x_n)^K  \\= \mathscr{M}_1  \frac{1}{\Delta^2} \sum_{j=-M}^M r_j x_1 \big( (m+j) \Delta \big)   \\
    k_1 (x_3(m\Delta) - x_2(m\Delta) ) + ... + k_k (x_3(m\Delta)-x_2(m\Delta))^K - k_1 (x_2(m\Delta) - x_1(m\Delta) )  - ... -k_K (x_2 (m\Delta)-x_1(m\Delta))^K \\= \mathscr{M}_2  \frac{1}{\Delta^2} \sum_{j=-M}^M r_j x_2 \big( (m+j) \Delta \big) \\
    \vdots \\
   k_1 (x_1(m\Delta)-x_n(m\Delta)) + ... + k_K (x_1(m\Delta)-x_n(m\Delta))^K - k_1 (x_n(m\Delta) - x_{n-1}(m\Delta))  - ... - k_K (x_n(m\Delta)-x_{n-1}(m\Delta))^K  \\= \mathscr{M}_n  \frac{1}{\Delta^2} \sum_{j=-M}^M r_j x_n \big( (m+j) \Delta \big).
\end{cases}
\label{12}
\end{align}
As the same structure holds for all $m=0,1,2,...,N$, we obtain a system of nonlinear algebraic equations. Given that the initial condition $x_1(0),x_2(0),...,x_n(0)$ is known, each displacement, for example, $x_1$, induces $N$ further variables $x_1(\Delta),..., x_1(N\Delta) $, so in total we have $ nM$ variables. Let $\xbf = (x_1,x_2,...,x_n)^T$ The outcome of our quantum Newton's method is then a block encoding of 
$$\rm diag \Big( \xbf(\Delta), \xbf(2\Delta), ..., \xbf(N\Delta) \Big)^T$$
which is equal to $\sum_{m=1}^N \ket{m-1}\bra{m-1} \otimes \rm diag \big(\xbf(m\Delta)\big)^T$ for which we can use the same procedure as in previous discussion to find out the potential energy of the system at any time $m\Delta$ between $(0,\mathscr{T})$. \\

\noindent
\textbf{Towards studying chaos in quantum regime.} Although the above 1-d spring-mass system features a simple physical model, however, its dynamics exhibit more complication than it seems to be, especially for $K>2$. More generally, understanding the pattern and behavior of dynamical systems over a long time period is a subject of great significance, namely, chaos theory. An important and popular property within chaos theory is called the Lyapunov exponent, which measures how quickly trajectories in the phase space diverge or converge, subject to initial conditions. Let $\xbf = (x_1,x_2,...,x_n)^T$, and suppose that we have a set of equations that governs the dynamic of some system in the first-order ordinary differential equation form: 
\begin{align}
\frac{d \xbf}{dt} = F(\xbf) \leftrightarrow  \begin{cases}
    &x_1' = f_1(x_1,x_2, ..., x_n) \\
    &x_2' = f_2(x_1, x_2,..., x_n) \\
    &\vdots \\
    & x_n' = f_n(x_1,x_2,....,x_n) 
\end{cases}
\end{align}
We divide the time interval $[0,\mathscr{T}]$ for example, $[0,\Delta, 2\Delta, ..., j\Delta, ..., N\Delta \equiv \mathscr{T}]$. Upon a discretization, for example,  $x_1'(j\Delta) = \frac{x_1( (j+1) \Delta)  - x_1((j-1)\Delta)}{2\Delta }$, and an initial condition $\xbf_0 =\big( x_1(0),x_2(0),...,x_n(0) \big)^T$, we obtain a nonlinear algebraic systems (similar to how we got Eqn.~\ref{12}). Application of quantum Newton's method yields the block encoding of $\rm diag \Big( \xbf(\Delta), \xbf(2\Delta), ..., \xbf(N\Delta) \Big)^T$, which is  $\sum_{m=1}^N \ket{m-1}\bra{m-1} \otimes \rm diag \big(\xbf(m\Delta)\big)^T $. Likewise, if we repeat the same procedure, but with different initial condition $\overline{\xbf_0} =\big( \overline{x_1}(0), \overline{x_2}(0),...,\overline{x_n}(0)\big)^T$, we obtain the block encoding of $\sum_{m=1}^N \ket{m-1}\bra{m -1} \otimes \rm diag \big(\overline{\xbf}(m\Delta) \big)^T$. To estimate the Lyapunov exponent, we divide $[0,\mathscr{T}]$ into $N_L$ renormalization steps, each with $T_L$ renormalization interval, i.e., $[0,\mathscr{T}] \longrightarrow [0, T_L, 2T_L,..., N_L T_L = \mathscr{T}]$. It is convenient to choose $T_L$ as some multiplication of $\Delta$ and hence $N_L \sim N$ (but being significantly smaller). For simplicity, at each step of renormalization $k T_L$, we define $\xbf(k T_L) = \big(x_1(kT_L),x_2(kT_L),...,x_n(k T_L)\big)^T$, and $\overline{\xbf(j\Delta)}$ is defined similarly. Let $d_k = || \overline{\xbf(kT_L)} -\xbf(k T_L)  || $. The Lyapunov exponent is calculated as $\lambda = \frac{1}{N_L T_L} \sum_{k=1}^{N_L} \ln \frac{d_k}{d_0}$ where $d_0 =|| \xbf_0  - \overline{\xbf_0}||$. Suppose that $k T_L = m\Delta$. Lemma \ref{lemma: sumencoding} allows us to construct the block encoding of
\begin{align}
    \frac{1}{2}\Big( \sum_{m=1}^N \ket{m-1}\bra{m -1} \otimes \rm diag \big(\xbf(m\Delta)\big)^T-  \sum_{m=1}^N \ket{m-1 }\bra{m -1} \otimes \rm diag \big(\overline{\xbf}(m\Delta) \big)^T   \Big) 
\end{align}
which is essentially $ \frac{1}{2} \Big( \sum_{m=1}^N \ket{m-1}\bra{m-1} \otimes \rm diag \big( \xbf(m\Delta)- \overline{\xbf}(m\Delta) \big)^T    \Big) $. We further note that $\rm diag \big( \xbf(m\Delta)- \overline{\xbf}(m\Delta) \big)^T  = \sum_{i=1}^n \ket{i-1}\bra{i-1}\otimes \Big( x_i(m\Delta) - \overline{x}_i(m\Delta)  \Big)  $, by applying the above unitary block encoding to the state $\ket{m}\frac{1}{\sqrt{n}} \sum_{i=1}^n \ket{i-1} $, according to definition \ref{def: blockencode}, we obtain the state $\ket{\bf 0} \sum_{i=1}^n \frac{1}{\sqrt{n}} \big( x_i(m\Delta) - \overline{x}_i(m\Delta)   \big) \ket{i-1} + \ket{\rm Garbage} $. Amplitude estimation allows us to reveal the value of $\frac{1}{\sqrt{n}} \sqrt{\big( x_i(m\Delta) - \overline{x}_i(m\Delta)   \big)^2}  = \frac{1}{\sqrt{n}} d_k $  (as $m\Delta = k T_L$). Given that $d_0 = || \xbf_0 - \overline{\xbf_0}|| $ is known, we can estimate the value of $\frac{d_k}{\sqrt{n} d_0}$, which can infer the value of $ \ln \frac{d_k}{d_0}$. Repeating the same procedure for all renormalization interval $0,T_L,2T_L,..,kT_L,.., N_L T_L$, we can then estimate the Lyapunov exponent $\lambda = \frac{1}{N_L T_L} \sum_{k=1}^{N_L} \ln \frac{d_k}{d_0}$. As a final remark, despite the equations \ref{eqn: newtonequation} that governs the dynamic of nonlinear coupled masses are not in the form of first-order ODE, however, it can be reduce to such a form by introducing another variable, namely, momentum $p_1 = \mathscr{M}_1 x_1', p_2 = \mathscr{M}_1 x_2',..., p_n = \mathscr{M}_n x_n'$. Given this, the right-hand side of Eqn.~\ref{eqn: newtonequation} is $p_i'$ and by definition, $p_i =\mathscr{M}_i x_i' $, for $i=1,2,...,n$. Hence, they form a first-order ordinary differential equations. The procedure we outlined above can thus be applied to investigate the chaotic behavior of the system, as indicated via the Lyapunov exponent. 

\noindent
\textbf{An improved quantum partial differential equation solver.} Although it is quite of independent interest, we point out a somewhat unexpected consequence of the above application. In Eqn.~\ref{eqn: circulant}, we encountered a circulant matrix. As shall be discussed in the Appendix \ref{sec: circulant}, the block encoding of the given circulant matrix can be constructed  logarithmically in the dimension $n$, which can be combined with result of \cite{gilyen2019quantum, low2017optimal,low2019hamiltonian} to directly obtain the inversion $\sim C^{-1}$ with a further logarithmical cost. In the Ref.~\cite{childs2017quantum} where the authors proposed high-precision quantum algorithm for partial differential equations, they also encountered a circulant matrix of size $n\times n$, however, they use the result of \cite{schuch2003programmable} with complexity being $\rm poly(n)$, to construct the exponent $\exp(-iCt)$ first before inverting it using technique of \cite{childs2017quantum}. Thus, our technique can be leveraged to construct an improved quantum partial differential equations solver, building upon that of \cite{childs2017quantum}. More details of \cite{childs2017quantum} and ours will be provided in the appendix \ref{sec: circulant} -- the result turns out to be that our improved version is faster by a factor of $n$.   

\noindent
\textbf{Motion planning and collision detection.} Robotics is garnering great attention due to its enormous practical impact. An important problem in robotics is motion planning, where a robot needs to be programmed so as to avoid obstacle on its path. The following figure is an illustration of motion planning. Four robots are designed to follow four paths, defined by some algebraic varieties $f_1(\xbf) = 0,f_2(\xbf) = 0, f_3(\xbf) = 0, f_4(\xbf) = 0$
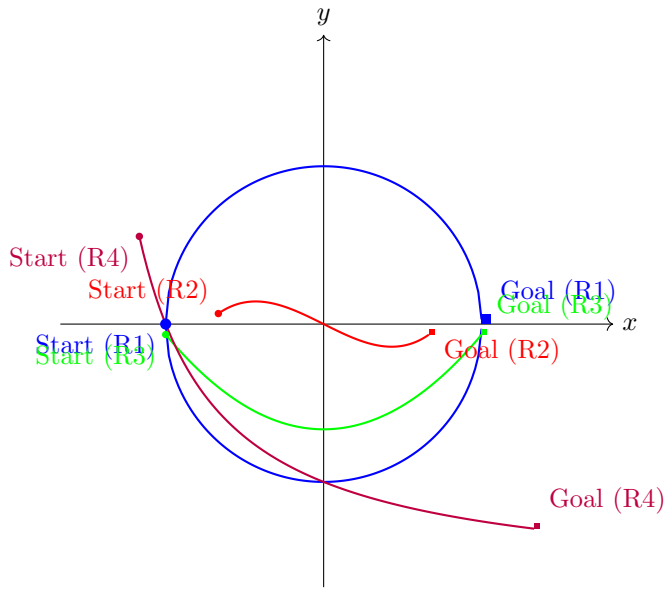
\begin{center}
    \begin{tikzpicture}[scale = 0.7]
        \draw[->] (0,5) -- (10.5,5) node[right] {$x$};
        \draw[->] (5,0) -- (5,10.5) node[above] {$y$};

        \draw[blue, thick, domain=2:8, samples=100] 
            plot (\x, {5 + sqrt(9 - (\x - 5)^2)});
        \draw[blue, thick, domain=2:8, samples=100] 
            plot (\x, {5 - sqrt(9 - (\x - 5)^2)});
        
        \draw[red, thick, domain=3:7, samples=100] 
            plot (\x, {0.1*(\x - 5)^3 - 0.5*(\x - 5) + 5});
        
        \draw[green, thick, domain=2:8, samples=100] 
            plot (\x, {0.2*(\x - 5)^2 + 3});
        
        \draw[purple, thick, domain=1.5:9, samples=100] 
            plot (\x, {10/\x});

        \fill[blue] (2, {5 + sqrt(9 - (2 - 5)^2)}) circle (3pt) node[below left] {Start (R1)};
        \fill[blue] (8, {5 + sqrt(9 - (8 - 5)^2)}) rectangle ++(5pt,5pt) node[above right] {Goal (R1)};
        
        \fill[red] (3, {0.1*(3 - 5)^3 - 0.5*(3 - 5) + 5}) circle (2pt) node[above left] {Start (R2)};
        \fill[red] (7, {0.1*(7 - 5)^3 - 0.5*(7 - 5) + 5}) rectangle ++(3pt,3pt) node[below right] {Goal (R2)};
        
        \fill[green] (2, {0.2*(2 - 5)^2 + 3}) circle (2pt) node[below left] {Start (R3)};
        \fill[green] (8, {0.2*(8 - 5)^2 + 3}) rectangle ++(3pt,3pt) node[above right] {Goal (R3)};
        
        \fill[purple] (1.5, {10/1.5}) circle (2pt) node[below left] {Start (R4)};
        \fill[purple] (9, {10/9}) rectangle ++(3pt,3pt) node[above right] {Goal (R4)};
        
    \end{tikzpicture}
\end{center}
As many robots are designed to move in a joint space, it is important for their trajectories not to cross each other, i.e., we want to avoid collision. Suppose we have two robots and we want to see if their paths would cross or not, we can look at the varieties $f_1(\xbf) =0 $ and $f_2(\xbf) = 0$. If there is a crossing point, it means that solution to the nonlinear systems exist. Applying our quantum Newton's method to this problem allows us to find the intersection. However, in reality, we might not be interested in the intersection, but simply want to see if there is an intersection, which signals the potential collision. Then we can instead look at  $g(\xbf) = f_1(\xbf) - f_2(\xbf)$ and see whether or not there exists $\xbf$ such that $g(\xbf) = 0$. This problem can be solved by our earlier result \ref{thm: multivariateroot}, which allows us to  examine the existence of root within the given domain almost exponentially faster than any classical approach.

\section{Conclusion}
We have successfully developed a quantum algorithm for dissecting roots and solving nonlinear algebraic equations. By reformulating the root detection problem as a sign change detection problem, we demonstrated that a quantum computer can determine the existence of a root within a given domain almost exponentially more efficiently than a classical computer. Moreover, this speed-up is maintained in both univariate and multivariate cases. Encouraged by this success, we extended our exploration to solving systems of nonlinear algebraic equations. We achieved this by translating the classical Newton's method into the quantum regime, taking advantage of recent advances in quantum algorithmic frameworks \cite{gilyen2019quantum} as well as prior constructions \cite{nghiem2024simple, nghiem2025quantum}. Given that non-linearity is a fundamental characteristic of many complex phenomena, our quantum algorithm for solving non-linear systems has significant potential for a wide range of applications. We have highlighted several practical applications where our method can be applied directly, including solving dense linear systems, analyzing the dynamics of nonlinear coupled masses, and enhancing collision detection in robotic motion planning, a highly relevant and valuable real-world problem. In particular, in the context of coupled masses system, we have shown how to leverage the quantum output to estimate key physical properties, such as equilibrium energy and potential energy, in logarithmical time. Thus, as a whole, our results provide an end-to-end application of quantum algorithms, achieving these advances without requiring artificial access to classical data. Ultimately, our work builds on and extends the previous success of \cite{nghiem2024simple, nghiem2025quantum}, further reinforcing the prospect of quantum advantage without requiring coherent quantum access, showing that a quantum computer can offer practically meaningful values. This solidifies a new and promising direction for future research in quantum computation. To name a few, although our quantum linear solver achieves quadratics running time in the dimension, it might not be close to being optimal. Determining the optimal threshold is hence an interesting point. A quantum algorithm for dense system with complexity being $\sim \sqrt{n}$ (assuming constant spectral norm) was introduced in \cite{wossnig2018quantum}, however, they require a different type of oracle access to the entries of the underlying matrix. Thus, whether it is possible to extend or improve our technique, so as to achieve anything close to sublinear scaling in complexity, is completely open. 


\section*{Acknowledgement}
 We acknowledge support from Center for Distributed Quantum Processing, Stony Brook University.

\appendix
\section{Preliminaries}
\label{sec: prelim}
Here, we summarize the main recipes of our work, which mostly derived in the seminal QSVT work \cite{gilyen2019quantum}. We keep the statements brief and precise for simplicity, with their proofs/ constructions referred to in their original works.

\begin{definition}[Block Encoding Unitary]~\cite{low2017optimal, low2019hamiltonian, gilyen2019quantum}
\label{def: blockencode} 
Let $A$ be some Hermitian matrix of size $N \times N$ whose matrix norm $|A| < 1$. Let a unitary $U$ have the following form:
\begin{align*}
    U = \begin{pmatrix}
       A & \cdot \\
       \cdot & \cdot \\
    \end{pmatrix}.
\end{align*}
Then $U$ is said to be an exact block encoding of matrix $A$. Equivalently, we can write:
\begin{align*}
    U = \ket{ \bf{0}}\bra{ \bf{0}} \otimes A + \cdots
\end{align*}
where $\ket{\bf 0}$ refers to the ancilla system required for the block encoding purpose. In the case where the $U$ has the form 
$$ U  =  \ket{ \bf{0}}\bra{ \bf{0}} \otimes \Tilde{A} + \cdots $$
where $|| \Tilde{A} - A || \leq \epsilon$ (with $||.||$ being the matrix norm), then $U$ is said to be an $\epsilon$-approximated block encoding of $A$. Furthermore, the action of $U$ on some quantum state $\ket{\bf 0}\ket{\phi}$ is:
\begin{align}
    \label{eqn: action}
    U \ket{\bf 0}\ket{\phi} = \ket{\bf 0} A\ket{\phi} +  \ket{\rm Garbage},
\end{align}
where $\ket{\rm Garbage }$ is a redundant state that is orthogonal to $\ket{\bf 0} A\ket{\phi}$. 
\end{definition}

The above definition has multiple natural corollaries. First, an arbitrary unitary $U$ block encodes itself. Suppose $A$ is block encoded by some matrix $U$. Next, then $A$ can be block encoded in a larger matrix by simply adding any ancilla (supposed to have dimension $m$), then note that $\Ibb_m \otimes U$ contains $A$ in the top-left corner, which is block encoding of $A$ again by definition. Further, it is almost trivial to block encode identity matrix of any dimension. For instance, we consider $\sigma_z \otimes \Ibb_m$ (for any $m$), which contains $\Ibb_m$ in the top-left corner.

\begin{lemma}[\cite{gilyen2019quantum} \revise{Block Encoding of a Density Matrix}]
\label{lemma: improveddme}
Let $\rho = \Tr_A \ket{\Phi}\bra{\Phi}$, where $\rho \in \mathbb{H}_B$, $\ket{\Phi} \in  \mathbb{H}_A \otimes \mathbb{H}_B$. Given unitary $U$ that generates $\ket{\Phi}$ from $\ket{\bf 0}_A \otimes \ket{\bf 0}_B$, then there exists a highly efficient procedure that constructs an exact unitary block encoding of $\rho$ using $U$ and $U^\dagger$ a single time, respectively.
\end{lemma}

The proof of the above lemma is given in \cite{gilyen2019quantum} (see their Lemma 45). \\



\begin{lemma}[Block Encoding of Product of Two Matrices]
\label{lemma: product}
    Given the unitary block encoding of two matrices $A_1$ and $A_2$, then there exists an efficient procedure that constructs a unitary block encoding of $A_1 A_2$ using each block encoding of $A_1,A_2$ one time. 
\end{lemma}

\begin{lemma}[\cite{camps2020approximate} \revise{Block Encoding of a Tensor Product}]
\label{lemma: tensorproduct}
    Given the unitary block encoding $\{U_i\}_{i=1}^m$ of multiple operators $\{M_i\}_{i=1}^m$ (assumed to be exact encoding), then, there is a procedure that produces the unitary block encoding operator of $\bigotimes_{i=1}^m M_i$, which requires \revise{parallel single uses} of 
    $\{U_i\}_{i=1}^m$ and $\mathcal{O}(1)$ SWAP gates. 
\end{lemma}
The above lemma is a result in \cite{camps2020approximate}. 
\begin{lemma}[\revise{\cite{gilyen2019quantum} Block Encoding of a  Matrix}]
\label{lemma: As}
    Given oracle access to $s$-sparse matrix $A$ of dimension $n\times n$, then an $\epsilon$-approximated unitary block encoding of $A/s$ can be prepared with gate/time complexity $\mathcal{O}\Big(\log n + \log^{2.5}(\frac{s^2}{\epsilon})\Big).$
\end{lemma}
This is presented in~\cite{gilyen2019quantum} (see their Lemma 48), and one can also find a review of the construction in~\cite{childs2017lecture}. We remark further that the scaling factor $s$ in the above lemma can be reduced by the preamplification method with further complexity $\mathcal{O}({s})$~\cite{gilyen2019quantum}.

\begin{lemma}[\cite{gilyen2019quantum} Linear combination of block-encoded matrices]
    Given unitary block encoding of multiple operators $\{M_i\}_{i=1}^m$. Then, there is a procedure that produces a unitary block encoding operator of \,$\sum_{i=1}^m \pm M_i/m $ in complexity $\mathcal{O}(m)$, e.g., using block encoding of each operator $M_i$ a single time. 
    \label{lemma: sumencoding}
\end{lemma}

\begin{lemma}[Scaling Block encoding] 
\label{lemma: scale}
    Given a block encoding of some matrix $A$ (as in~\ref{def: blockencode}), then the block encoding of $A/p$ where $p > 1$ can be prepared with an extra $\mathcal{O}(1)$ cost.  
\end{lemma}
To show this, we note that the matrix representation of RY rotational gate is
\begin{align}
   R_Y(\theta) = \begin{pmatrix}
        \cos(\theta/2) & -\sin(\theta/2) \\
        \sin(\theta/2) & \cos(\theta/2) 
    \end{pmatrix}.
\end{align}
If we choose $\theta$ such that $\cos(\theta/2) = 1/p$, then Lemma~\ref{lemma: tensorproduct} allows us to construct block encoding of $R_Y(\theta) \otimes \mathbb{I}_{{\rm dim}(A)}$  (${\rm dim}(A)$ refers to dimension of matirx $A$), which contains the diagonal matrix of size ${\rm dim}(A) \times {\rm dim}(A)$ with entries $1/p$. Then Lemma~\ref{lemma: product} can construct block encoding of $(1/p) \ \mathbb{I}_{{\rm dim}(A)} \cdot A = A/p$.  \\

The following is called amplification technique:
\begin{lemma}[\cite{gilyen2019quantum} Theorem 30; \revise{\bf Amplification}]\label{lemma: amp_amp}
Let $U$, $\Pi$, $\widetilde{\Pi} \in {\rm End}(\mathcal{H}_U)$ be linear operators on $\mathcal{H}_U$ such that $U$ is a unitary, and $\Pi$, $\widetilde{\Pi}$ are orthogonal projectors. 
Let $\gamma>1$ and $\delta,\epsilon \in (0,\frac{1}{2})$. 
Suppose that $\widetilde{\Pi}U\Pi=W \Sigma V^\dagger=\sum_{i}\varsigma_i\ket{w_i}\bra{v_i}$ is a singular value decomposition. 
Then there is an $m= \mathcal{O} \Big(\frac{\gamma}{\delta}
\log \left(\frac{\gamma}{\epsilon} \right)\Big)$ and an efficiently computable $\Phi\in\mathbb{R}^m$ such that
\begin{equation}
\left(\bra{+}\otimes\widetilde{\Pi}_{\leq\frac{1-\delta}{\gamma}}\right)U_\Phi \left(\ket{+}\otimes\Pi_{\leq\frac{1-\delta}{\gamma}}\right)=\sum_{i\colon\varsigma_i\leq \frac{1-\delta}{\gamma} }\tilde{\varsigma}_i\ket{w_i}\bra{v_i} , \text{ where } \Big|\!\Big|\frac{\tilde{\varsigma}_i}{\gamma\varsigma_i}-1 \Big|\!\Big|\leq \epsilon.
\end{equation}
Moreover, $U_\Phi$ can be implemented using a single ancilla qubit with $m$ uses of $U$ and $U^\dagger$, $m$ uses of C$_\Pi$NOT and $m$ uses of C$_{\widetilde{\Pi}}$NOT gates and $m$ single qubit gates.
Here,
\begin{itemize}
\item C$_\Pi$NOT$:=X \otimes \Pi + I \otimes (I - \Pi)$ and a similar definition for C$_{\widetilde{\Pi}}$NOT; see Definition 2 in \cite{gilyen2019quantum},
\item $U_\Phi$: alternating phase modulation sequence; see Definition 15 in \cite{gilyen2019quantum},
\item $\Pi_{\leq \delta}$, $\widetilde{\Pi}_{\leq \delta}$: singular value threshold projectors; see Definition 24 in \cite{gilyen2019quantum}.
\end{itemize}
\end{lemma}

\begin{lemma}[Projector]
\label{lemma: projector}
The block encoding of a projector $\ket{j-1}\bra{j-1}$ (for any $j=1,2, ...,n$) by a circuit of depth $\mathcal{O}\big( \log n \big)$ 
\end{lemma}
\noindent
\textit{Proof.} First we note that it takes a circuit of depth $\mathcal{O}(1)$ to generate $\ket{j-1}$ from $\ket{0}$. Then Lemma \ref{lemma: improveddme} can be used to construct the block encoding of $\ket{j-1}\bra{j-1}$.

\section{Details of Section \ref{sec: blockencodingjacobian}    }
\label{sec: detailjacobian}
Here we provide full details on the algorithm for constructing the block encoding of Jacobian $\Jcal(\xbf)$ for a given $\xbf$, with some unitary circuit that generates $\rm diag(\xbf)$.

\begin{method}[Algorithm for constructing the block encoding of $\Jcal(\xbf)$]
    \label{method: algorithmjacobian}
\end{method}
\begin{enumerate}
    \item First we use Lemma \ref{lemma: tensorproduct} to construct the block encoding of $\ket{j-1}\bra{j-1} \otimes \rm diag  \big( \bigtriangledown \frac{1}{M}  f_j(\xbf)\big)$ for all $j$. 
    \item Use Lemma \ref{lemma: sumencoding} to construct the block encoding of $ \sum_{j=1}^n \frac{1}{n}\ket{j-1}\bra{j-1} \otimes \rm diag  \big( \bigtriangledown \frac{1}{M}  f_j(\xbf)\big)$, which is equal to:
    \begin{align}
         \frac{1}{n}  \sum_{j=1}^n \ket{j-1}\bra{j-1} \otimes  \sum_{k=1}^n\big( \bigtriangledown \frac{1}{M}  f_j(\xbf)_k \big)\ket{k-1}\bra{k-1}
    \end{align}
    \item Since the SWAP operators between two $\log n$-qubit register is already a unitary, we can use it with Lemma \ref{lemma: product} to transform the above block-encoded operator into:
    \begin{align}
         \frac{1}{n}  \sum_{j=1}^n \ket{j-1}\bra{j-1} \otimes  \sum_{k=1}^n\big(\frac{1}{M}  \bigtriangledown f_j(\xbf)_k \big)\ket{k-1}\bra{k-1} \longrightarrow  \frac{1}{n}  \sum_{j=1}^n \sum_{k=1}^n \ket{j-1}\bra{k-1} \otimes  \big( \frac{1}{M} \bigtriangledown f_j(\xbf)_k \big)\ket{k-1}\bra{j-1}
    \end{align}
    \item For a fixed $j$, we consider the term $\sum_{k=1}^n  \frac{1}{n} \ket{j-1}\bra{k-1} \otimes \big( \bigtriangledown f_j(\xbf)_k \big)\ket{k-1}\bra{j-1} $. Given that $H^{\otimes \log n} \otimes \Ibb$ is already a unitary, we use Lemma \ref{lemma: product} to construct the block encoding of $ \frac{1}{nM}\sum_{k=1}^n\Big(  \ket{j-1}\bra{k-1} \otimes \big( \bigtriangledown f_j(\xbf)_k \big)\ket{k-1}\bra{j-1} \Big) \cdot \Big(  H^{\otimes \log n} \otimes \Ibb\Big)  $. Note that $\bra{k-1}H^{\otimes \log n} = \frac{1}{\sqrt{n}} \big( \bra{0}+ \sum_{p=2}^n  (\pm) \bra{p-1}\big)$, so we obtain the block encoding of:
    \begin{align}
       \frac{1}{nM\sqrt{n}}\sum_{k=1}^n \ket{j-1} \Big( \bra{0}  \Big) \otimes \big( \bigtriangledown f_j(\xbf)_k \big)\ket{k-1}\bra{j-1} + \frac{1}{nM\sqrt{n}} \sum_{k=1}^n \sum_{p=2}^n \pm \ket{j-1} \bra{p-1}\otimes \big( \bigtriangledown f_j(\xbf)_k \big)\ket{k-1}\bra{j-1} 
    \end{align}
    where the $\pm$ comes from the expansion of $H^{\otimes \log n} \ket{k-1}$. 
    \item Consider specifically the first term of the above, and take into account of summation over $j$:
    \begin{align}
       \sum_{j=1}^n \frac{1}{Mn\sqrt{n}} \ket{j-1} \bra{0} \otimes \sum_{k=1}^n \big( \bigtriangledown   f_j(\xbf)_k \big) \ket{k-1}\bra{j-1} &=  \sum_{j=1}^n \frac{1}{Mn\sqrt{n}} \ket{j-1} \bra{0} \otimes \big( \bigtriangledown   f_j(\xbf) \big) \bra{j-1}
    \end{align}
    \item Now we use block encoding of $H^{\otimes \log n} \otimes \Ibb$ and Lemma \ref{lemma: product} to construct the block encoding of $ \Big( H^{\otimes \log n} \otimes \Ibb\Big) \cdot \Big( \sum_{j=1}^n \frac{1}{M n\sqrt{n}} \ket{j-1} \bra{0} \otimes  \bigtriangledown f_j(\xbf) \bra{j-1} \Big)$, which is
    \begin{align}
        &\frac{1}{M n^2} \ket{0}\bra{0} \otimes \sum_{j=1}^n \bigtriangledown f_j(\xbf)  \bra{j-1}  + \frac{1}{M n^2} \sum_{j'\neq 0} \sum_{j=1}^n\ket{j'} \bra{0} \otimes  \big( \bigtriangledown f_j(\xbf) \big) \bra{j-1} 
    \end{align}
    \item According to Definition \ref{def: blockencode}, the above operator is a block encoding of $\frac{1}{M n^2} \sum_{j=1}^n \bigtriangledown f_j(\xbf)  \bra{j-1} $, which is $\frac{1}{M n^2} \Jcal(\xbf)^T$. The transpose of such unitary is the block encoding of $\frac{1}{M n^2} \Jcal(\xbf)$.
    
    The complexity of the above algorithm is contributed mostly by $n$ usage of Lemma \ref{lemma: diagonalgradient} and a single use of SWAP operator between two $\log n$-qubit registers. Thus, the total complexity is $\mathcal{O}\big( n T_X  + \log n  \big)$. 
\end{enumerate}

\section{Details of Section \ref{sec: blockencodingF}}
\label{sec: continuation}
\noindent
\textbf{First approach.}  First we use Lemma \ref{lemma: product} to obtain the following transformation
\begin{align}
   \rm diag(\xbf) \longrightarrow \big( \rm diag(\xbf)\big)^i = \bigoplus_{j=1}^n x_j^i 
    \label{eqn: 12}
\end{align}
for $i=1,2,...,K$. Let $U_i$ denotes the unitary block encoding of the operator on the right-hand side of the above. According to Def.~\ref{def: blockencode} and Eqn.~\ref{eqn: action}, we have that $U_i \ket{\bf 0} \ket{a} = \ket{\bf 0} \sum_{k=1}^n x^i_k \sqrt{a_{i,k}} \ket{k-1} + \ket{\rm Garbage}$. Defining the resultant state as $\ket{\Phi_1} $ with $U_{\Phi_1} = U_i \cdot (\Ibb \otimes U_a)$ as a generating quantum circuit, i.e., $U_i \cdot (\Ibb \otimes U_a) \ket{0} = \ket{\Phi_1} $, and $\ket{\Phi_2} = \ket{\bf 0}\ket{a}$ with $U_{\Phi_2}=  \Ibb\otimes U_a$ as a generating quantum circuit. Consider the following state $ \frac{1}{\sqrt{2}} \ket{0} \ket{\Phi_1 } + \frac{1}{\sqrt{2}} \ket{1} \ket{\Phi_2}$, which can be generated by first creating $\frac{1}{\sqrt{2} } (\ket{0} + \ket{1}) \ket{\bf 0} $ and use $U_{\Phi_1}, U_{\Phi_2}$ controlled by $\ket{0},\ket{1}$ respectively to generate $\ket{\Phi_1},\ket{\Phi_2}$ entangled to corresponding register. Now we apply Hadamard gate to the first register, and append an ancilla $\ket{0}$, then we obtain $\frac{1}{2} \ket{0}\ket{0} \big( \ket{\Phi_1 }  + \ket{\Phi_2} \big) + \frac{1}{2} \ket{0}\ket{1}\big(  \ket{\Phi_1 }  - \ket{\Phi_2}\big) $. Use the second qubit as controlled bit, and apply $X$ on the first ancilla qubit, we obtain the state:
\begin{align}
    \frac{1}{2} \ket{0}\ket{0} \big( \ket{\Phi_1 }  + \ket{\Phi_2} \big) + \frac{1}{2} \ket{1}\ket{1}\big(  \ket{\Phi_1 }  - \ket{\Phi_2}\big) 
    \label{16}
\end{align}
Tracing out the second and last register, we have the following density state on the first ancilla $\rho = \frac{1}{2} \big(1 + \braket{\Phi_1,\Phi_2}\big) \ket{0}\bra{0} + \frac{1}{2} \big( 1 - \braket{\Phi_1,\Phi_2}  \big) \ket{1}\bra{1}$. We note that Lemma \ref{lemma: improveddme} allows us to block-encode $\rho$. It is trivial to obtain the block encoding of $\frac{1}{2} \ket{0}\bra{0} + \frac{1}{2}\ket{1}\bra{1}$, e.g., use Lemma \ref{lemma: scale} with scaling factor $p=2$ combined with a block encoding of $\Ibb_2 = \ket{0}\bra{0} + \ket{1}\bra{1}$, which is trivial to prepare. Then we use Lemma \ref{lemma: sumencoding} to construct the block encoding of $\rho - \frac{1}{2} \ket{0}\bra{0} - \frac{1}{2}\ket{1}\bra{1} $, which is $\frac{\braket{\Phi_1,\Phi_2}}{4} \ket{0}\bra{0 } - \frac{\braket{\Phi_1,\Phi_2}}{4} \ket{1}\bra{1} $

We further note that $\braket{\Phi_1,\Phi_2} = \bra{\bf 0}\bra{a} \Big(   \ket{\bf 0} \sum_{k=1}^n x^i_k \sqrt{a_{i,k}} \ket{k-1} + \ket{\rm Garbage}\Big)$. By using the orthogonal relation between $\ket{\bf 0}\ket{a}$ and $\ket{\rm Garbage} $, we have such overlaps turn out to be equal to:
\begin{align}
     \sum_{k=1}^n x^i_k a_{i,k}  = a_{i,1} x_1^i + a_{i,2} x_2^i + \cdots + a_{i,n} x_n^i
\end{align}
Thus, we have achieved the block encoding of $\frac{1}{4} (a_{i,1} x_1^i + a_{i,2} x_2^i + \cdots + a_{i,n} x_n^i) \ket{0}\bra{0}$ for a specific $i$. Repeat the same procedure for all other $i$, and use Lemma \ref{lemma: sumencoding} to construct the block encoding of $\frac{1}{4K} \Big(\sum_{i=1}^K a_{i,1} x_1^i + a_{i,2} x_2^i + \cdots + a_{i,n} x_n^i\Big) \ket{0}\bra{0}$. The factor $4K$ can be removed using Lemma \ref{lemma: amp_amp}, resulting a block encoding of $ \Big(\sum_{i=1}^K a_{i,1} x_1^i + a_{i,2} x_2^i + \cdots + a_{i,n} x_n^i \Big)\ket{0}\bra{0} = f(\xbf) \ket{0}\bra{0}$. To transform it into the block encoding of $\ket{j}\bra{j}\otimes f(\xbf)$, we simply need to use a block encoding of $\ket{j-1}\bra{j-1}$ (which is simple to generate as we mentioned in previous section), of $f(\xbf) \ket{0}\bra{0}$ combined with Lemma \ref{lemma: tensorproduct}. 

The remaining instances of $f(\xbf)$ can be constructed using the above procedure in a straightforward manner. More specifically, for $f(\xbf) =\sum_{i=1}^K  \big( a_{i,1} x_1 + a_{i,2} x_2 + \cdots + a_{i,n} x_n \big)^i  $. We can use a similar procedure as above (e.g., starting from Eqn.~\ref{eqn: 12} without raising the power $i$) to first construct the block encoding of $ \big(   a_{i,1} x_1 + a_{i,2} x_2 + \cdots + a_{i,n} x_n\big)  \ket{0}\bra{0}$, then use Lemma \ref{lemma: product} to transform it into $ \big(   a_{i,1} x_1 + a_{i,2} x_2 + \cdots + a_{i,n} x_n\big)^i  \ket{0}\bra{0}$. Then Lemma \ref{lemma: sumencoding} can be used to construct the block encoding of $  \frac{1}{4K} \Big(\sum_{i=1}^K a_{i,1} x_1 + a_{i,2} x_2 + \cdots + a_{i,n} x_n\Big)^i \ket{0}\bra{0}$, with the factor $4K$ subsequently removed with Lemma \ref{lemma: amp_amp}. Finally we use Lemma \ref{lemma: tensorproduct} to construct the block encoding of $\ket{j-1}\bra{j-1} \otimes \Big(\sum_{i=1}^K a_{i,1} x_1 + a_{i,2} x_2 + \cdots + a_{i,n} x_n\Big)^i $.  

For the next case, $f(\xbf) = \prod_{i=1}^K  \big( a_{i,1} x_1 + a_{i,2}x_2 + \cdots + a_{i,n}x_n  + b_i \big)^i $, we can use the result above -- the block encoding of $\big(   a_{i,1} x_1 + a_{i,2} x_2 + \cdots + a_{i,n} x_n\big)^i  \ket{0}\bra{0} $ for $i=1,2,...,K$, combined with Lemma \ref{lemma: product} to construct the block encoding of their products, then we obtain the block encoding of $ \prod_{i=1}^K  \big( a_{i,1} x_1 + a_{i,2}x_2 + \cdots + a_{i,n}x_n  + b_i \big)^i \ket{0}\bra{0} $. A simple usage of Lemma \ref{lemma: tensorproduct} yields the block encoding of $ \Big(  \prod_{i=1}^K  \big( a_{i,1} x_1 + a_{i,2}x_2 + \cdots + a_{i,n}x_n  + b_i \big)^i \Big) \ket{j-1}\bra{j-1}$ for any $j=1,2,...,n$. 

We have completed all cases. Now for all $j=1,2,...,n$, each $f_j(\xbf)$ belongs to one of those types discussed above, so we can use the same procedure to construct the block encoding of $\ket{0}\bra{0} \otimes f_1(\xbf), \ket{1}\bra{1}\otimes f_2(\xbf)$, ..., $\ket{n-1}\bra{n-1} \otimes f_n(\xbf)$ -- each of which is constructed with complexity $\mathcal{O}\big( T_X + \log n \big)$ where $\mathcal{O}(\log n)$ comes from the application of Lemma \ref{lemma: improveddme} to block encode $\rho$, see below Eqn.~\ref{16}. Then Lemma \ref{lemma: sumencoding} can be applied to construct the block encoding of $\frac{1}{n} \sum_{j=1}^n \ket{j-1}\bra{j-1}\otimes f_j(\xbf) = \frac{1}{n}\rm diag \big( \Fsr(\xbf) \big)$. Assume that the depth of $U_a$ is $\mathcal{O}(1)$, and the depth of the unitary that contains $\rm diag (\xbf)$ is $T_X$ as defined earlier. The complexity of this step is $\mathcal{O}\big( n T_X + n \log n \big)$. \\

\noindent
\textbf{Second approach. } Now we provide a second approach, showing that if all $\{ f_j(\xbf)\}_{i=1}^n$ of 
$$\Fsr(\xbf) = \big(f_1(\xbf),f_2(\xbf),...,f_n(\xbf) \big)^T$$ 
are sharing the same form, then the complexity of producing $\rm diag \Fsr(\xbf)$ can be further improved.  Recall that in the main text, we have supposed that each $f_j(\xbf)$ can be either of $\sum_{i=1}^K a_{i,1} x_1^i + a_{i,2} x_2^i + \cdots + a_{i,n} x_n^i, \   \sum_{i=1}^K  \big( a_{i,1} x_1 + a_{i,2} x_2 + \cdots + a_{i,n} x_n \big)^i, \  \prod_{i=1}^K  \big( a_{i,1} x_1 + a_{i,2}x_2 + \cdots + a_{i,n}x_n  + b_i \big)^i,  $. Additionally, there is a unitary that generates a state $\sim ( \sqrt{a_{i,1}}, \sqrt{a_{i,2}}, ..., \sqrt{a_{i,n}})$. The strategy in Section \ref{sec: blockencodingF} relies on building the block encoding of $\ket{0}\bra{0} f_1(\xbf), \ket{1}\bra{1} f_2(\xbf), ..., \ket{n-1}\bra{n-1} f_n(\xbf)$, from which the block encoding of $ \sum_{j=1}^n \frac{1}{n} f_j(\xbf) \ket{j-1}\bra{j-1}$ can be constructed through the Lemma \ref{lemma: sumencoding}. This step incurs the complexity scaling $\mathcal{O}(n)$ -- which is quite insufficient and it turns out that there is a possible improvement. Let all $f_j(\xbf)$ having the same form as:
\begin{align}
    \Fsr(\xbf) = \begin{cases}
        f_1(\xbf) \\
        f_2(\xbf) \\
        \vdots \\
        f_n(\xbf)
    \end{cases} = \begin{cases}
        \sum_{i=1}^K a_{i,1}^1 x_1^i + a_{i,2}^1 x_2^i + ... + a_{i,n}^1 x_n^i \\
         \sum_{i=1}^K a_{i,1}^2 x_1^i + a_{i,2}^2 x_2^i + ... + a_{i,n}^2 x_n^i \\
          \vdots \\
        \sum_{i=1}^K a_{i,1}^n x_1^i + a_{i,2}^n x_2^i + ... + a_{i,n}^n x_n^i \\
    \end{cases}
\end{align}
Now if we are provided with a unitary $U_i$ that generates the state 
\begin{align}
    \ket{\Phi}_i &=  \big( \sqrt{a^1_{i,1}}, \sqrt{a^1_{i,2}}, ..., \sqrt{a^1_{i,n}}, \sqrt{a^2_{i,1}}, \sqrt{a^2_{i,2}}, ..., \sqrt{a^2_{i,n}}, ..., \sqrt{a^n_{i,1}}, \sqrt{a^n_{i,2}}, ..., \sqrt{a^n_{i,n}} \big)^T  \\
    &=  \sum_{k=1}^n \ket{k-1} \otimes \big(  \sqrt{a_{i,1}^k}, \sqrt{a_{i,2}^k}, ... ,\sqrt{a_{i,n}^k} \big)^T
\end{align}

From the block encoding of $\rm diag (\xbf) = \bigoplus_{j=1}^n x_j$, we first use Lemma \ref{lemma: product} to construct the block encoding of $\bigoplus_{j=1}^n x_j^i $. Then we use Lemma \ref{lemma: sumencoding} to construct the block encoding of $\frac{1}{2}\Big( \Ibb_n +\bigoplus_{j=1}^n x_j^i  \Big) = \bigoplus_{j=1}^n \frac{1}{2}(1+ x_j^i)$. The reason for this step is to make sure that the resultant operator is positive, which is important in order to apply the following result from \cite{gilyen2019quantum, chakraborty2018power}:
\begin{lemma}[Positive Power Exponent \cite{gilyen2019quantum},\cite{chakraborty2018power}]
\label{lemma: positive}
    Given an $\delta$-approximated block encoding of a positive matrix $O$ such that 
    $$ \frac{\Ibb}{\kappa} \leq O \leq \Ibb. $$
   Let $\delta = \epsilon/ (\kappa \log^3(\frac{\kappa}{\epsilon})) $ and $c \in (0,1)$. Then we can implement an $\epsilon$-approximated block encoding of $M^c/2$ in time complexity $\mathcal{O}( \kappa T_O \log^2 (\frac{\kappa}{\epsilon})  )$, where $T_O$ is the complexity of block encoding of $O$. 
\end{lemma}
The above lemma can be used to construct the block encoding of $\bigoplus_{j=1}^n \sqrt{ \frac{1}{2}(1+x_j^i)}$. Given that block encoding of $\Ibb_n$ is simple to prepare, we can use it with Lemma \ref{lemma: tensorproduct} to construct the block encoding of $ \Ibb_n \otimes  \bigoplus_{j=1}^n \sqrt{\frac{1}{2}(1+x_j^i)} = \sum_{k=1}^n \ket{k-1}\bra{k-1}\otimes \sqrt{\frac{1}{2}(1+x_j^i)}  $. Denote such unitary as $U_X$. We have:
\begin{align}
  U_X \ket{\bf 0}\ket{\Phi}_i  &= \ket{\bf 0} \Big( \Ibb_n \otimes  \bigoplus_{k=1}^n \sqrt{\frac{1}{2}(1+x_j^i)}\Big) \ket{A} + \ket{\rm Garbage} \\
  &= \ket{\bf 0} \sum_{k=1}^n \ket{k-1} \Big(   \sqrt{a_{i,1}^k \frac{1}{2}(1+x_1^i)} \ket{0} + \sqrt{a_{i,2}^k \frac{1}{2}(1+x_2^i) } \ket{1} + \sqrt{a_{i,3}^k \frac{1}{2}(1+x_3^i)} \ket{2}+ \\&  ... + \sqrt{a_{i,n}^k \frac{1}{2}(1+x_n^i) } \ket{n-1}   \Big) + \ket{\rm Garbage}
\end{align}
Now we append another $\log (n)$ qubit ancilla initialized in $\ket{0}$, and use the register $\ket{k}$ as controlled bit to apply $X$ gates on the ancilla, we obtain:
\begin{align}
    \ket{\bf 0} \sum_{k=1}^n \ket{k-1} \ket{k-1} \Big(   \sqrt{a_{i,1}^k\frac{1}{2}(1+ x_1^i)} \ket{0} + \sqrt{a_{i,2}^k \frac{1}{2}(1+x_2^i)} \ket{1} + \\ \sqrt{a_{i,3}^k \frac{1}{2} (1+x_3^i)} \ket{2}+ ... + \sqrt{a_{i,n}^k \frac{1}{2}(1+x_n^i) } \ket{n-1}   \Big) +  \ket{\rm Garbage'}\ket{\rm Garbage}
\end{align}
Tracing out the last two register, we obtain a density state 
\begin{align}
    \sigma = \ket{\bf 0}\bra{\bf 0} \otimes  \sum_{k=1}^n \ket{k-1}\bra{k-1} \big(  a^k_{i,1} \frac{1}{2}(1+x_1^i) + a^k_{i,2} \frac{1}{2}(1+x_2^i) + ... + a_{i,n}^k \frac{1}{2}(1+ x_n^i)  \big) + (...) 
\end{align}
where $(...)$ denotes irrelevant terms. Lemma \ref{lemma: improveddme} allows us to block encode $\sigma$, which is also a block encoding of $ \sum_{k=1}^n \ket{k-1}\bra{k-1} \big(  a^k_{i,1} \frac{1}{2}(1+x_1^i )+ a^k_{i,2} \frac{1}{2}(1+x_2^i) + ... + a_{i,n}^k \frac{1}{2}(1+x_n^i)  \big)$ according to Definition \ref{def: blockencode}. Using the same procedure, it is straightforward to show that the block encoding of 
\begin{align}
    \sum_{k=1}^n \ket{k-1}\bra{k-1} \frac{1}{2}\big( a^k_{i,1} +  a^k_{i,2} + ... + a_{i,n}^k\big)
\end{align}
can be constructed. Then we use Lemma \ref{lemma: sumencoding} to construct the block encoding of: 
\begin{align}
   \frac{1}{2}\left( \sum_{k=1}^n \ket{k-1}\bra{k-1} \big(  a^k_{i,1} \frac{1}{2}(1+x_1^i )+ a^k_{i,2} \frac{1}{2}(1+x_2^i) + ... + a_{i,n}^k \frac{1}{2}(1+x_n^i)  \big)- \sum_{k=1}^n \ket{k-1}\bra{k-1} \frac{1}{2}\big( a^k_{i,1} +  a^k_{i,2} + ... + a_{i,n}^k\big) \right)
\end{align}
which is exactly $  \frac{1}{4}\sum_{k=1}^n \ket{k-1}\bra{k-1} \big(  a_{i,1}^k + a_{i,2}^k + ... + a_{i,n}^k \big) $. Now we construct the same block encoding but for all $i=1,2,..,K$. Then we use Lemma \ref{lemma: sumencoding} to construct the block encoding of $\sum_{k=1}^n \ket{k-1}\bra{k -1} \frac{1}{4K}\sum_{i=1}^K \Big(  a^k_{i,1}x_1^i + a^k_{i,2}x_2^i + ... + a_{i,n}^k x_n^i \Big) $. The factor $4K$ can be removed using Lemma \ref{lemma: amp_amp}, so we obtain the block encoding of $\sum_{k=1}^n \ket{k-1}\bra{k-1} f_k(\xbf) $. Let $K = \mathcal{O}(1)$ (independent of $n$). The complexity of this approach mainly comes from the usage of Lemma \ref{lemma: improveddme} and the complexity of producing $\rm diag(\xbf$, and also $\ket{A_i}$ for all $i=1,2,...,K$. Assuming efficient generation of $\ket{A_i}$, which means the depth of $U_{A_i}$  is at most $\mathcal{O}( \log n)$, thus the total complexity is $\mathcal{O}\Big( \log n+ T_X  \Big)$. \\

We remind that in the main text, section \ref{sec: blockencodingF}, we pointed out that if all the functions within $\Fsr(\xbf)$ shares similar form, then the block encoding of operator $\frac{1}{M} \bigoplus_{j=1}^n \rm diag\big( \bigtriangledown  f_j(\xbf) \big)$ can be constructed with complexity $\mathcal{O}\big( T_X + \log n\big)$. In the following, by extending the technique used above, we show it explicitly. For the system having the form as above, e.g.,
\begin{align}
    \Fsr(\xbf) = \begin{cases}
        f_1(\xbf) \\
        f_2(\xbf) \\
        \vdots \\
        f_n(\xbf)
    \end{cases} = \begin{cases}
        \sum_{i=1}^K a_{i,1}^1 x_1^i + a_{i,2}^1 x_2^i + ... + a_{i,n}^1 x_n^i \\
         \sum_{i=1}^K a_{i,1}^2 x_1^i + a_{i,2}^2 x_2^i + ... + a_{i,n}^2 x_n^i \\
          \vdots \\
        \sum_{i=1}^K a_{i,1}^n x_1^i + a_{i,2}^n x_2^i + ... + a_{i,n}^n x_n^i \\
    \end{cases}
\end{align}
The gradient of $f_1(\xbf)$ is apparently: 
\begin{align}
    \bigtriangledown f_1(\xbf) =\sum_{i=1}^K \begin{pmatrix}
        a_{i,1}^1 i x_1^{i-1} \\
        a_{i,2}^1 i x_2^{i-1} \\
        \vdots \\
        a_{i,n}^1 i x_n^{i-1}
    \end{pmatrix}
\end{align}
Now for a given $i$, we use the block encoding of $\rm diag(\xbf)$ and raise it to power $i-1$, then we have $\rm diag \big( x_1^{i-1}, x_2^{i-1}, ...,x_n^{i-1}  \big)^T$. This block encoding is then used to construct the block encoding of $\Ibb_n \otimes \rm diag \big( x_1^{i-1}, x_2^{i-1}, ...,x_n^{i-1}  \big)^T$.  Recall that for each $i$, we have $U_i$ that generates the state $\ket{\Phi}_i = \sum_{k=1}^n \ket{k-1} \otimes \big(  \sqrt{a_{i,1}^k}, \sqrt{a_{i,2}^k}, ... ,\sqrt{a_{i,n}^k} \big)^T$. Lemma \ref{lemma: diagonal} allows us to construct the block encoding of $ \bigoplus_{k=1}^n  \rm diag \big( \sqrt{a_{i,1}^k}, \sqrt{a_{i,2}^k}, ... ,\sqrt{a_{i,n}^k}  \big)^T $. Then we use Lemma \ref{lemma: product} to obtain the block encoding of $\bigoplus_{k=1}^n  \rm diag \big( a_{i,1}^k, a_{i,2}^k, ... , a_{i,n}^k \big)^T  $. Now we use Lemma \ref{lemma: product} and Lemma \ref{lemma: scale} to construct the block encoding of: 
\begin{align}
    \Ibb_n \otimes \rm diag \big( x_1^{i-1}, x_2^{i-1}, ...,x_n^{i-1}  \big)^T \cdot 
 \frac{i}{M}\bigoplus_{k=1}^n  \rm diag \big( a_{i,1}^k, a_{i,2}^k, ... , a_{i,n}^k \big)^T  
\end{align}
Repeating the same procedure for all $i=1,2,...,K$, and use Lemma \ref{lemma: sumencoding}, we can construct the block encoding of:
\begin{align}
    \frac{1}{K} \sum_{i=1}^K \Ibb_n \otimes \rm diag \big( x_1^{i-1}, x_2^{i-1}, ...,x_n^{i-1}  \big)^T \cdot 
 \frac{i}{M}\bigoplus_{k=1}^n  \rm diag \big( a_{i,1}^k, a_{i,2}^k, ... , a_{i,n}^k \big)^T
\end{align}
By a simple algebraic procedure, it can be shown that the above operator is exactly $\frac{1}{M} \bigoplus_{j=1}^n \rm diag \big( \bigtriangledown f_j(\xbf)  \big)$. The above procedure use the block encoding of $\rm diag (\xbf)$ $\mathcal{O}(1)$ times, and Lemma \ref{lemma: diagonal} $\mathcal{O}(1)$ times, thus the total complexity is $\mathcal{O}( T_X + \log n)$.

For the remaining types, $f_j(\xbf)$ can be either $ \sum_{i=1}^K  \big( a_{i,1} x_1 + a_{i,2} x_2 + \cdots + a_{i,n} x_n \big)^i, \  \prod_{i=1}^K  \big( a_{i,1} x_1 + a_{i,2}x_2 + \cdots + a_{i,n}x_n  + b_i \big)^i$, the technique above can be extended in a straightforward way, and hence, procedure as well as complexity is essentially the same.

\section{Circulant matrix and implementation}
\label{sec: circulant}
In the following we further elaborate on the properties of circulant matrix and how to implement it. A circulant matrix $C$ of size $n\times n$ is a special type of Toeplitz matrix, that is formally defined as:
\begin{align}
    C = \begin{pmatrix}
        c_1 & c_2 & \cdots & c_n \\
        c_n & c_1 & \cdots & c_{n-1} \\
        \vdots & \vdots & \ddots & \vdots \\
        c_2 & c_3 & \cdots & c_1
    \end{pmatrix}
\end{align}
That is, the $i$-th row is the $i-1$-th row shifted to the right by one step. Let $\lambda_1,\lambda_2,...,\lambda_n$ be the eigenvalues of $C$. For any $k=1,2,...,n$,  we have:
\begin{align}
    \lambda_k = \sum_{j=1}^n c_j \omega^{j\cdot k}
\end{align}
where $\omega = \exp( -\frac{2\pi i}{n})$ is the primitive $n$-th root of unity. A special property of circulant matrix $C$, which was also employed in \cite{childs2021high}, is that $C$ can be diagonalized by quantum Fourier transform $F_n$ (QFT) on $\log (n)$ qubits. It means that $\Lambda = F_n^\dagger C F_n$ is diagonal matrix with entries being eigenvalues of $C$ as defined above, i.e.,
\begin{align}
  \Lambda = \begin{pmatrix}
      \lambda_1 &  0 & \cdots & 0 \\
      0 & \lambda_2 & \cdots & 0 \\
      0 & 0 & \ddots & 0 \\
      0 & 0 & 0 & \lambda_n
  \end{pmatrix}
\end{align}
Suppose that all $c_1,c_2,...,c_n$ are known, and apparently  the method in \cite{zhang2022quantum} can be used to construct the quantum state $\sim (\lambda_1,\lambda_2,...,\lambda_n)$ (where again we remind that $\sim$ refers to a possible multiplication/normalization factor). Then Lemma \ref{lemma: diagonal} allows us to construct the block encoding of $\sim \rm diag \ (\lambda_1,\lambda_2,...,\lambda_n)^T = \Lambda$ (if there is a multiplication factor, we can use Lemma \ref{lemma: amp_amp} to remove it). Since QFT has known efficient implementation as a unitary circuit, we can use it with Lemma \ref{lemma: product} to construct the block encoding of $F_n \Lambda F_n^\dagger $, which yields the block encoding of $C$. The complexity of the aforementioned procedure is contributed mostly by an application of Lemma \ref{lemma: diagonal} and two QFT circuits, thus resulting in total complexity $\mathcal{O}\big( \log n \big)$. 

In the main text, we have pointed out that the above construction can be used to improve the quantum partial differential equation solver introduced in \cite{childs2021high}. In the following we elaborate more details on such an improvement. Consider, for example, the 2-d Poisson equation on the domain $\mathscr{D} = [-1,1]^2$:
\begin{align}
    \frac{\partial^2 f(x,y,z)}{\partial^2 x} + \frac{\partial^2 f(x,y,z)}{\partial^2 y }  = g(x,y)
\end{align}
The method in \cite{childs2021high} is built upon the discretization of the above equations, using central finite difference formula (which we did use in the main text) $f''(x) \approx \frac{1}{\Delta^2} \sum_{j=-M}^M r_j f(x + j \Delta)$ with the coefficients:
\begin{align}
    r_j = \begin{cases}
        \frac{2(-1)^{j+1} (M!)^2}{ j^2 (M-j)! (M+j)!}  \ \text{for } \  j \in [1,2,...,M] \\
        - 2 \sum_{j=1}^M r_j \ \text{for} \ j= 0\\
        r_{-j} \ \text{for} \ j \in -[1,2,...,M]
    \end{cases}
\end{align}
By discretizing the domain $\mathscr{D}$ into 2-d rectangular lattice and using the above formula, they obtain the following equations:
\begin{align}
    \frac{1}{\Delta^2} C \Vec{u} = \Vec{g}
\end{align}
where $\Vec{u}$ is a vector that stores the values of $f(x,y)$ on the grid, and $\Vec{g}$ stores the value of $g(x,y)$ on the boundary of $[-1,1]^2$, and in particular, $C$ is circulant with definition and related properties defined above. In order to solve for $\Vec{u}$, we need to invert $L$, which was done in \cite{childs2021high} as follows. First they use the method in \cite{schuch2003programmable} to construct $\exp(-i \Lambda t)$. Then they use QFT to construct $F_n \exp (-i \Lambda t) F_n^\dagger$, which is exactly $\exp(-i C t)$. Such exponentiation of $C$ allows them to apply the technique in \cite{childs2017quantum} to construct $C^{-1}$ and apply it to $\Vec{g}$ (assume to have efficient preparation), thus completing the quantum PDE solving algorithm. We point out two things: the conditional number of $C$ is $\mathcal{O}\big(n^2\big)$, as proved in \cite{childs2017quantum}; and the cost of simulating $\exp(- i \Lambda t)$ is $\mathcal{O}(n)$. Thus, the total complexity of \cite{childs2021high} is $\mathcal{O}(n^3)$ (see their Theorem 1 for more explicit information). On the other hand, by using our method to first construct (the block encoding of) $C$ and directly invert it using \cite{gilyen2019quantum}, the complexity scaling is $\mathcal{O}(n^2 \log n)$ -- which is faster by almost a factor of $n$ compared to \cite{childs2017quantum}.

\bibliography{ref.bib}{}

\begin{thebibliography}{10}

\bibitem{manin1980computable}
Yuri Manin.
\newblock Computable and uncomputable.
\newblock {\em Sovetskoye Radio, Moscow}, 128:15, 1980.

\bibitem{benioff1980computer}
Paul Benioff.
\newblock The computer as a physical system: A microscopic quantum mechanical hamiltonian model of computers as represented by turing machines.
\newblock {\em Journal of statistical physics}, 22:563--591, 1980.

\bibitem{shor1999polynomial}
Peter~W Shor.
\newblock Polynomial-time algorithms for prime factorization and discrete logarithms on a quantum computer.
\newblock {\em SIAM review}, 41(2):303--332, 1999.

\bibitem{grover1996fast}
Lov~K Grover.
\newblock A fast quantum mechanical algorithm for database search.
\newblock In {\em Proceedings of the twenty-eighth annual ACM symposium on Theory of computing}, pages 212--219, 1996.

\bibitem{deutsch1985quantum}
David Deutsch.
\newblock Quantum theory, the church--turing principle and the universal quantum computer.
\newblock {\em Proceedings of the Royal Society of London. A. Mathematical and Physical Sciences}, 400(1818):97--117, 1985.

\bibitem{deutsch1992rapid}
David Deutsch and Richard Jozsa.
\newblock Rapid solution of problems by quantum computation.
\newblock {\em Proceedings of the Royal Society of London. Series A: Mathematical and Physical Sciences}, 439(1907):553--558, 1992.

\bibitem{feynman2018simulating}
Richard~P Feynman.
\newblock Simulating physics with computers.
\newblock In {\em Feynman and computation}, pages 133--153. CRC Press, 2018.

\bibitem{lloyd1996universal}
Seth Lloyd.
\newblock Universal quantum simulators.
\newblock {\em Science}, 273(5278):1073--1078, 1996.

\bibitem{berry2007efficient}
Dominic~W Berry, Graeme Ahokas, Richard Cleve, and Barry~C Sanders.
\newblock Efficient quantum algorithms for simulating sparse hamiltonians.
\newblock {\em Communications in Mathematical Physics}, 270(2):359--371, 2007.

\bibitem{berry2012black}
Dominic~W Berry and Andrew~M Childs.
\newblock Black-box hamiltonian simulation and unitary implementation.
\newblock {\em Quantum Information and Computation}, 12:29--62, 2009.

\bibitem{berry2014high}
Dominic~W Berry.
\newblock High-order quantum algorithm for solving linear differential equations.
\newblock {\em Journal of Physics A: Mathematical and Theoretical}, 47(10):105301, 2014.

\bibitem{berry2015hamiltonian}
Dominic~W Berry, Andrew~M Childs, and Robin Kothari.
\newblock Hamiltonian simulation with nearly optimal dependence on all parameters.
\newblock In {\em 2015 IEEE 56th annual symposium on foundations of computer science}, pages 792--809. IEEE, 2015.

\bibitem{low2017optimal}
Guang~Hao Low and Isaac~L Chuang.
\newblock Optimal hamiltonian simulation by quantum signal processing.
\newblock {\em Physical review letters}, 118(1):010501, 2017.

\bibitem{low2019hamiltonian}
Guang~Hao Low and Isaac~L Chuang.
\newblock Hamiltonian simulation by qubitization.
\newblock {\em Quantum}, 3:163, 2019.

\bibitem{childs2022quantum}
Andrew~M Childs, Jiaqi Leng, Tongyang Li, Jin-Peng Liu, and Chenyi Zhang.
\newblock Quantum simulation of real-space dynamics.
\newblock {\em Quantum}, 6:860, 2022.

\bibitem{o2016scalable}
Peter~JJ O’Malley, Ryan Babbush, Ian~D Kivlichan, Jonathan Romero, Jarrod~R McClean, Rami Barends, Julian Kelly, Pedram Roushan, Andrew Tranter, Nan Ding, et~al.
\newblock Scalable quantum simulation of molecular energies.
\newblock {\em Physical Review X}, 6(3):031007, 2016.

\bibitem{cerezo2021variational}
Marco Cerezo, Andrew Arrasmith, Ryan Babbush, Simon~C Benjamin, Suguru Endo, Keisuke Fujii, Jarrod~R McClean, Kosuke Mitarai, Xiao Yuan, Lukasz Cincio, et~al.
\newblock Variational quantum algorithms.
\newblock {\em Nature Reviews Physics}, 3(9):625--644, 2021.

\bibitem{babbush2023exponential}
Ryan Babbush, Dominic~W Berry, Robin Kothari, Rolando~D Somma, and Nathan Wiebe.
\newblock Exponential quantum speedup in simulating coupled classical oscillators.
\newblock {\em Physical Review X}, 13(4):041041, 2023.

\bibitem{babbush2018low}
Ryan Babbush, Nathan Wiebe, Jarrod McClean, James McClain, Hartmut Neven, and Garnet Kin-Lic Chan.
\newblock Low-depth quantum simulation of materials.
\newblock {\em Physical Review X}, 8(1):011044, 2018.

\bibitem{mitarai2023perturbation}
Kosuke Mitarai, Kiichiro Toyoizumi, and Wataru Mizukami.
\newblock Perturbation theory with quantum signal processing.
\newblock {\em Quantum}, 7:1000, 2023.

\bibitem{robert2021resource}
Anton Robert, Panagiotis~Kl Barkoutsos, Stefan Woerner, and Ivano Tavernelli.
\newblock Resource-efficient quantum algorithm for protein folding.
\newblock {\em npj Quantum Information}, 7(1):38, 2021.

\bibitem{kitaev1995quantum}
A~Yu Kitaev.
\newblock Quantum measurements and the abelian stabilizer problem.
\newblock {\em arXiv preprint quant-ph/9511026}, 1995.

\bibitem{aharonov2006polynomial}
Dorit Aharonov, Vaughan Jones, and Zeph Landau.
\newblock A polynomial quantum algorithm for approximating the jones polynomial.
\newblock In {\em Proceedings of the thirty-eighth annual ACM symposium on Theory of computing}, pages 427--436, 2006.

\bibitem{childs2010relationship}
Andrew~M Childs.
\newblock On the relationship between continuous-and discrete-time quantum walk.
\newblock {\em Communications in Mathematical Physics}, 294(2):581--603, 2010.

\bibitem{childs2021high}
Andrew~M Childs, Jin-Peng Liu, and Aaron Ostrander.
\newblock High-precision quantum algorithms for partial differential equations.
\newblock {\em Quantum}, 5:574, 2021.

\bibitem{childs2017quantum}
Andrew~M Childs, Robin Kothari, and Rolando~D Somma.
\newblock Quantum algorithm for systems of linear equations with exponentially improved dependence on precision.
\newblock {\em SIAM Journal on Computing}, 46(6):1920--1950, 2017.

\bibitem{childs2021quantum}
Andrew~M Childs, Shih-Han Hung, and Tongyang Li.
\newblock Quantum query complexity with matrix-vector products.
\newblock {\em arXiv preprint arXiv:2102.11349}, 2021.

\bibitem{lloyd2013quantum}
Seth Lloyd, Masoud Mohseni, and Patrick Rebentrost.
\newblock Quantum algorithms for supervised and unsupervised machine learning.
\newblock {\em arXiv preprint arXiv:1307.0411}, 2013.

\bibitem{lloyd2014quantum}
Seth Lloyd, Masoud Mohseni, and Patrick Rebentrost.
\newblock Quantum principal component analysis.
\newblock {\em Nature Physics}, 10(9):631--633, 2014.

\bibitem{lloyd2016quantum}
Seth Lloyd, Silvano Garnerone, and Paolo Zanardi.
\newblock Quantum algorithms for topological and geometric analysis of data.
\newblock {\em Nature communications}, 7(1):1--7, 2016.

\bibitem{lloyd2020quantum}
Seth Lloyd, Maria Schuld, Aroosa Ijaz, Josh Izaac, and Nathan Killoran.
\newblock Quantum embeddings for machine learning.
\newblock {\em arXiv preprint arXiv:2001.03622}, 2020.

\bibitem{hauke2020perspectives}
Philipp Hauke, Helmut~G Katzgraber, Wolfgang Lechner, Hidetoshi Nishimori, and William~D Oliver.
\newblock Perspectives of quantum annealing: Methods and implementations.
\newblock {\em Reports on Progress in Physics}, 83(5):054401, 2020.

\bibitem{liu2021efficient}
Jin-Peng Liu, Herman~{\O}ie Kolden, Hari~K Krovi, Nuno~F Loureiro, Konstantina Trivisa, and Andrew~M Childs.
\newblock Efficient quantum algorithm for dissipative nonlinear differential equations.
\newblock {\em Proceedings of the National Academy of Sciences}, 118(35):e2026805118, 2021.

\bibitem{liu2018quantum}
Nana Liu and Patrick Rebentrost.
\newblock Quantum machine learning for quantum anomaly detection.
\newblock {\em Physical Review A}, 97(4):042315, 2018.

\bibitem{liu2024towards}
Junyu Liu, Minzhao Liu, Jin-Peng Liu, Ziyu Ye, Yunfei Wang, Yuri Alexeev, Jens Eisert, and Liang Jiang.
\newblock Towards provably efficient quantum algorithms for large-scale machine-learning models.
\newblock {\em Nature Communications}, 15(1):434, 2024.

\bibitem{durr1996quantum}
Christoph Durr and Peter Hoyer.
\newblock A quantum algorithm for finding the minimum.
\newblock {\em arXiv preprint quant-ph/9607014}, 1996.

\bibitem{bauer2020quantum}
Bela Bauer, Sergey Bravyi, Mario Motta, and Garnet Kin-Lic Chan.
\newblock Quantum algorithms for quantum chemistry and quantum materials science.
\newblock {\em Chemical Reviews}, 120(22):12685--12717, 2020.

\bibitem{brassard1997quantum}
Gilles Brassard, Peter Hoyer, and Alain Tapp.
\newblock Quantum algorithm for the collision problem.
\newblock {\em arXiv preprint quant-ph/9705002}, 1997.

\bibitem{brassard2002quantum}
Gilles Brassard, Peter Hoyer, Michele Mosca, and Alain Tapp.
\newblock Quantum amplitude amplification and estimation.
\newblock {\em Contemporary Mathematics}, 305:53--74, 2002.

\bibitem{jordan2012quantum}
Stephen~P Jordan, Keith~SM Lee, and John Preskill.
\newblock Quantum algorithms for quantum field theories.
\newblock {\em Science}, 336(6085):1130--1133, 2012.

\bibitem{leyton2008quantum}
Sarah~K Leyton and Tobias~J Osborne.
\newblock A quantum algorithm to solve nonlinear differential equations.
\newblock {\em arXiv preprint arXiv:0812.4423}, 2008.

\bibitem{nachman2021quantum}
Benjamin Nachman, Davide Provasoli, Wibe~A De~Jong, and Christian~W Bauer.
\newblock Quantum algorithm for high energy physics simulations.
\newblock {\em Physical review letters}, 126(6):062001, 2021.

\bibitem{jordan2005fast}
Stephen~P Jordan.
\newblock Fast quantum algorithm for numerical gradient estimation.
\newblock {\em Physical review letters}, 95(5):050501, 2005.

\bibitem{garnerone2012adiabatic}
Silvano Garnerone, Paolo Zanardi, and Daniel~A Lidar.
\newblock Adiabatic quantum algorithm for search engine ranking.
\newblock {\em Physical review letters}, 108(23):230506, 2012.

\bibitem{gilyen2022quantum}
Andr{\'a}s Gily{\'e}n, Seth Lloyd, Iman Marvian, Yihui Quek, and Mark~M Wilde.
\newblock Quantum algorithm for petz recovery channels and pretty good measurements.
\newblock {\em Physical Review Letters}, 128(22):220502, 2022.

\bibitem{miessen2023quantum}
Alexander Miessen, Pauline~J Ollitrault, Francesco Tacchino, and Ivano Tavernelli.
\newblock Quantum algorithms for quantum dynamics.
\newblock {\em Nature Computational Science}, 3(1):25--37, 2023.

\bibitem{haah2021quantum}
Jeongwan Haah, Matthew~B Hastings, Robin Kothari, and Guang~Hao Low.
\newblock Quantum algorithm for simulating real time evolution of lattice hamiltonians.
\newblock {\em SIAM Journal on Computing}, 52(6):FOCS18--250, 2021.

\bibitem{arrazola2019quantum}
Juan~Miguel Arrazola, Timjan Kalajdzievski, Christian Weedbrook, and Seth Lloyd.
\newblock Quantum algorithm for nonhomogeneous linear partial differential equations.
\newblock {\em Physical Review A}, 100(3):032306, 2019.

\bibitem{kerenidis2019quantum}
Iordanis Kerenidis, Jonas Landman, and Anupam Prakash.
\newblock Quantum algorithms for deep convolutional neural networks.
\newblock {\em arXiv preprint arXiv:1911.01117}, 2019.

\bibitem{hallgren2007polynomial}
Sean Hallgren.
\newblock Polynomial-time quantum algorithms for pell's equation and the principal ideal problem.
\newblock {\em Journal of the ACM (JACM)}, 54(1):1--19, 2007.

\bibitem{tang2018quantum}
Ewin Tang.
\newblock Quantum-inspired classical algorithms for principal component analysis and supervised clustering.
\newblock {\em arXiv preprint arXiv:1811.00414}, 4, 2018.

\bibitem{tang2019quantum}
Ewin Tang.
\newblock A quantum-inspired classical algorithm for recommendation systems.
\newblock In {\em Proceedings of the 51st annual ACM SIGACT symposium on theory of computing}, pages 217--228, 2019.

\bibitem{tang2021quantum}
Ewin Tang.
\newblock Quantum principal component analysis only achieves an exponential speedup because of its state preparation assumptions.
\newblock {\em Physical Review Letters}, 127(6):060503, 2021.

\bibitem{aaronson2015read}
Scott Aaronson.
\newblock Read the fine print.
\newblock {\em Nature Physics}, 11(4):291--293, 2015.

\bibitem{bravyi2018quantum}
Sergey Bravyi, David Gosset, and Robert K{\"o}nig.
\newblock Quantum advantage with shallow circuits.
\newblock {\em Science}, 362(6412):308--311, 2018.

\bibitem{bravyi2020quantum}
Sergey Bravyi, David Gosset, Robert K{\"o}nig, and Marco Tomamichel.
\newblock Quantum advantage with noisy shallow circuits.
\newblock {\em Nature Physics}, 16(10):1040--1045, 2020.

\bibitem{maslov2021quantum}
Dmitri Maslov, Jin-Sung Kim, Sergey Bravyi, Theodore~J Yoder, and Sarah Sheldon.
\newblock Quantum advantage for computations with limited space.
\newblock {\em Nature Physics}, 17(8):894--897, 2021.

\bibitem{liu2021rigorous}
Yunchao Liu, Srinivasan Arunachalam, and Kristan Temme.
\newblock A rigorous and robust quantum speed-up in supervised machine learning.
\newblock {\em Nature Physics}, 17(9):1013--1017, 2021.

\bibitem{nghiem2024simple}
Nhat~A Nghiem.
\newblock Simple quantum gradient descent without coherent oracle access.
\newblock {\em arXiv preprint arXiv:2412.18309}, 2024.

\bibitem{nghiem2025quantum}
Nhat~A Nghiem.
\newblock Quantum computer does not need coherent quantum access for advantage.
\newblock {\em arXiv preprint arXiv:2503.02515}, 2025.

\bibitem{gilyen2019quantum}
Andr{\'a}s Gily{\'e}n, Yuan Su, Guang~Hao Low, and Nathan Wiebe.
\newblock Quantum singular value transformation and beyond: exponential improvements for quantum matrix arithmetics.
\newblock In {\em Proceedings of the 51st Annual ACM SIGACT Symposium on Theory of Computing}, pages 193--204, 2019.

\bibitem{camps2020approximate}
Daan Camps and Roel Van~Beeumen.
\newblock Approximate quantum circuit synthesis using block encodings.
\newblock {\em Physical Review A}, 102(5):052411, 2020.

\bibitem{grover2000synthesis}
Lov~K Grover.
\newblock Synthesis of quantum superpositions by quantum computation.
\newblock {\em Physical review letters}, 85(6):1334, 2000.

\bibitem{grover2002creating}
Lov Grover and Terry Rudolph.
\newblock Creating superpositions that correspond to efficiently integrable probability distributions.
\newblock {\em arXiv preprint quant-ph/0208112}, 2002.

\bibitem{plesch2011quantum}
Martin Plesch and {\v{C}}aslav Brukner.
\newblock Quantum-state preparation with universal gate decompositions.
\newblock {\em Physical Review A}, 83(3):032302, 2011.

\bibitem{schuld2018supervised}
Maria Schuld and Francesco Petruccione.
\newblock {\em Supervised learning with quantum computers}, volume~17.
\newblock Springer, 2018.

\bibitem{nakaji2022approximate}
Kouhei Nakaji, Shumpei Uno, Yohichi Suzuki, Rudy Raymond, Tamiya Onodera, Tomoki Tanaka, Hiroyuki Tezuka, Naoki Mitsuda, and Naoki Yamamoto.
\newblock Approximate amplitude encoding in shallow parameterized quantum circuits and its application to financial market indicators.
\newblock {\em Physical Review Research}, 4(2):023136, 2022.

\bibitem{marin2023quantum}
Gabriel Marin-Sanchez, Javier Gonzalez-Conde, and Mikel Sanz.
\newblock Quantum algorithms for approximate function loading.
\newblock {\em Physical Review Research}, 5(3):033114, 2023.

\bibitem{zoufal2019quantum}
Christa Zoufal, Aur{\'e}lien Lucchi, and Stefan Woerner.
\newblock Quantum generative adversarial networks for learning and loading random distributions.
\newblock {\em npj Quantum Information}, 5(1):103, 2019.

\bibitem{qian2019quantum}
Peng Qian, Wei-Cong Huang, and Gui-Lu Long.
\newblock A quantum algorithm for solving systems of nonlinear algebraic equations.
\newblock {\em arXiv preprint arXiv:1903.05608}, 2019.

\bibitem{xue2021quantum}
Cheng Xue, Yuchun Wu, and Guoping Guo.
\newblock Quantum newton’s method for solving the system of nonlinear equations.
\newblock In {\em Spin}, volume~11, page 2140004. World Scientific, 2021.

\bibitem{xue2022quantum}
Cheng Xue, Xiao-Fan Xu, Yu-Chun Wu, and Guo-Ping Guo.
\newblock Quantum algorithm for solving a quadratic nonlinear system of equations.
\newblock {\em Physical Review A}, 106(3):032427, 2022.

\bibitem{nghiem2024quantumnonlinear}
Nhat~A Nghiem and Tzu-Chieh Wei.
\newblock Quantum algorithm for solving nonlinear algebraic equations.
\newblock {\em arXiv preprint arXiv:2404.03810}, 2024.

\bibitem{zhang2022quantum}
Xiao-Ming Zhang, Tongyang Li, and Xiao Yuan.
\newblock Quantum state preparation with optimal circuit depth: Implementations and applications.
\newblock {\em Physical Review Letters}, 129(23):230504, 2022.

\bibitem{burton2009newton}
T.~A. Burton.
\newblock Newton's method and differential equations.
\newblock {\em Journal of Mathematical Analysis and Applications}, 350(2):850--862, 2009.

\bibitem{pacsca2011formal}
L.~Pacska and Others.
\newblock Formal methods in differential equations.
\newblock {\em Annals of Mathematics and Artificial Intelligence}, 63(4):451--470, 2011.

\bibitem{rattew2023non}
Arthur~G Rattew and Patrick Rebentrost.
\newblock Non-linear transformations of quantum amplitudes: Exponential improvement, generalization, and applications.
\newblock {\em arXiv preprint arXiv:2309.09839}, 2023.

\bibitem{guo2024nonlinear}
Naixu Guo, Kosuke Mitarai, and Keisuke Fujii.
\newblock Nonlinear transformation of complex amplitudes via quantum singular value transformation.
\newblock {\em Physical Review Research}, 6(4):043227, 2024.

\bibitem{nghiem2022quantum}
Nhat~A Nghiem and Tzu-Chieh Wei.
\newblock Quantum algorithm for estimating eigenvalue.
\newblock {\em arXiv preprint arXiv:2211.06179}, 2022.

\bibitem{nghiem2024improved}
Nhat~A Nghiem, Hiroki Sukeno, Shuyu Zhang, and Tzu-Chieh Wei.
\newblock Improved quantum power method and numerical integration using quantum singular value transformation.
\newblock {\em arXiv preprint arXiv:2407.11744}, 2024.

\bibitem{nghiem2023improved}
Nhat~A Nghiem and Tzu-Chieh Wei.
\newblock Improved quantum algorithms for eigenvalues finding and gradient descent.
\newblock {\em arXiv preprint arXiv:2312.14786}, 2023.

\bibitem{harrow2009quantum}
Aram~W Harrow, Avinatan Hassidim, and Seth Lloyd.
\newblock Quantum algorithm for linear systems of equations.
\newblock {\em Physical review letters}, 103(15):150502, 2009.

\bibitem{wossnig2018quantum}
Leonard Wossnig, Zhikuan Zhao, and Anupam Prakash.
\newblock Quantum linear system algorithm for dense matrices.
\newblock {\em Physical review letters}, 120(5):050502, 2018.

\bibitem{nghiem2025new}
Nhat~A Nghiem.
\newblock New quantum algorithm for principal component analysis.
\newblock {\em arXiv preprint arXiv:2501.07891}, 2025.

\bibitem{huang2019near}
Hsin-Yuan Huang, Kishor Bharti, and Patrick Rebentrost.
\newblock Near-term quantum algorithms for linear systems of equations.
\newblock {\em arXiv preprint arXiv:1909.07344}, 2019.

\bibitem{schuch2003programmable}
Norbert Schuch and Jens Siewert.
\newblock Programmable networks for quantum algorithms.
\newblock {\em Physical review letters}, 91(2):027902, 2003.

\bibitem{childs2017lecture}
Andrew~M Childs.
\newblock Lecture notes on quantum algorithms.
\newblock {\em Lecture notes at University of Maryland}, 2017.

\bibitem{chakraborty2018power}
Shantanav Chakraborty, Andr{\'a}s Gily{\'e}n, and Stacey Jeffery.
\newblock The power of block-encoded matrix powers: improved regression techniques via faster hamiltonian simulation.
\newblock {\em arXiv preprint arXiv:1804.01973}, 2018.

\end{thebibliography}
\bibliographystyle{unsrt}

\clearpage
\newpage
\onecolumngrid

\end{document}